\documentclass[lettersize, journal]{IEEEtran}

\usepackage{cite}
\usepackage{amsmath,amssymb,amsfonts}
\usepackage{algorithm}
\usepackage{algorithmic}
\usepackage{graphicx}
\usepackage{textcomp}
\usepackage{xcolor}
\usepackage{balance}
\usepackage{siunitx}
\usepackage{caption}
\usepackage{subcaption}
\usepackage{multirow}
\usepackage{makecell}
\usepackage{optidef}
\usepackage{multirow}
\usepackage{booktabs}
\usepackage{amsthm} 

\usepackage[acronym]{glossaries}
\usepackage{subcaption} 

\makeglossaries
\newacronym{isac}{ISAC}{Integrated Sensing and Communication}
\newacronym{snr}{SNR}{Signal-to-Noise Ratio}
\newacronym{bs}{BS}{base station}
\newacronym{ue}{UE}{user equipment}
\newacronym{csi}{CSI}{Channel State Information}
\newacronym{ofdm}{OFDM}{Orthogonal Frequency Division Multiplexing}
\newacronym{los}{LoS}{Line of sight}
\newacronym{nlos}{nLoS}{non-line of sight}
\newacronym{fm}{FM}{Foundation Model}
\newacronym{wmfm}{WMFM}{Wireless Multimodal Foundation Model}
\newacronym{mlp}{MLP}{multi-layer perceptron}
\newacronym{infonce}{InfoNCE}{Information Noise-Contrastive Estimation}
\newacronym{e2e}{E2E}{end-to-end}
\newacronym{tsne}{t-SNE}{t-distributed Stochastic Neighbor Embedding}
\newacronym{cnn}{CNN}{convolutional neural network}
\newacronym{mi}{MI}{mutual information}
\newacronym{lmm}{LMM}{Large Multi-modal Model}
\newacronym{ai}{AI}{Artificial Intelligence}
\newacronym{nlp}{NLP}{Natural Language Processing}
\newacronym{lwm}{LWM}{Large Wireless Model}
\newacronym{6g}{6G}{6th Generation}
\newacronym{ssl}{SSL}{self-supervised learning}
\newacronym{rgb}{RGB}{red-green-blue}
\newtheorem{theorem}{Theorem}

\begin{document}

\title{Wireless Multimodal Foundation Model (WMFM): Integrating Vision and Communication Modalities for 6G ISAC Systems \vspace{0ex}}

\author{
        Mohammad Farzanullah,
        Han~Zhang,
        Akram Bin Sediq,
        Ali Afana,
        and~Melike~Erol-Kantarci,~\IEEEmembership{Fellow, IEEE}
\thanks{Mohammad Farzanullah, Han Zhang (deceased), and Melike Erol-Kantarci are with the School of Electrical Engineering and Computer Science, University of Ottawa, Ottawa, ON K1N 6N5, Canada (e-mail: mfarz086@uottawa.ca; melike.erolkantarci@uottawa.ca).}
\thanks{Akram Bin Sediq and Ali Afana are with Ericsson, Ottawa, K2K 2V6, Canada (e-mail:
akram.bin.sediq@ericsson.com; ali.afana@ericsson.com)
}
}
\maketitle
\vspace{-2em} 
\begin{abstract}

The emergence of multimodal foundation models has revolutionized learning paradigms by enabling joint understanding across diverse data types. In the context of next-generation wireless networks, integrating sensing and communication modalities presents a unique opportunity to develop generalizable and data-efficient models. In this work, we introduce the contrastive learning based Wireless Multimodal Foundation Model (WMFM), a large-scale framework that jointly learns from wireless channel coefficients and visual imagery. The WMFM is pretrained using contrastive learning, a self-supervised learning technique that aligns embeddings of camera and channel data without requiring explicit labels. The pretrained encoders are then frozen and employed as feature extractors, with lightweight task-specific heads, fine-tuned for downstream tasks, including user localization and LoS/nLoS classification. Extensive experiments on the DeepVerse6G dataset demonstrate that the proposed WMFM achieves a 17\% improvement in balanced accuracy for LoS/nLoS classification and a 48.5\% reduction in localization error compared to the end-to-end (E2E) benchmark, while reducing training time by up to 90-fold. 
Even when trained with as little as 20\% of the data, the WMFM-based heads outperform the fully supervised E2E model, underscoring their robustness and data-efficient learning.
The proposed approach establishes a foundation for scalable, multimodal learning in \gls{isac} systems, paving the way for intelligent and adaptive 6G networks.
\end{abstract}

\begin{IEEEkeywords}
Foundation models, self-supervised learning, contrastive learning, multimodal learning, 6G networks, integrated sensing and communication
\end{IEEEkeywords}
\glsresetall

\section{Introduction}

\IEEEPARstart{T}{he} integration of \gls{ai} into future \gls{6g} networks is anticipated to fundamentally transform wireless systems by enabling intelligent, adaptive, and data-driven operation across all layers of the communication stack \cite{elsayed2019ai}. 
Unlike previous generations that rely on hand-crafted signal processing pipelines, \gls{6g} envisions an \gls{ai}-native design where key functionalities, such as beamforming, channel prediction, or semantic communication are directly learned from data. 
This paradigm shift calls for models that are generalizable, data-efficient, and capable of transferring knowledge across heterogeneous wireless tasks.
Recent advances in \glspl{fm} offer a promising solution. \glspl{fm} such as GPT, Claude, Gemini, and DALL-E are large-scale models pretrained on broad and diverse datasets, and have demonstrated strong zero-shot and few-shot capabilities across many domains. In contrast, traditional wireless machine learning approaches often adopt task-specific discriminative \gls{ai}, requiring separate models and large labeled datasets for each application, leading to substantial data collection and engineering costs \cite{shoaib2024convergence}.

In contrast, a \gls{fm} can be pretrained on large, heterogeneous wireless datasets, potentially using unlabeled or weakly labeled data, and then fine-tuned for multiple tasks, significantly reducing the need for task-specific retraining. Most importantly, generative \gls{ai} models offer the ability to capture underlying data distributions and structures, making them especially valuable in wireless contexts where real-world data are 
often noisy, incomplete, and, expensive to acquire
\cite{celik2024dawn}.

Such models enable cross-task knowledge transfer and improved generalization by learning rich, task-agnostic representations that capture fundamental patterns in wireless data. These representations can be effectively adapted to a wide range of downstream applications. Recent works \cite{fm_survey1, yang2025generative} highlight the potential of a unified physical-layer \gls{fm} to model complex RF environments and serve as a reusable backbone across tasks. By leveraging large-scale pretraining, \glspl{fm} can accelerate the development of \gls{ai}-driven wireless systems, particularly in capturing propagation effects that are difficult to train with traditional discriminative approaches.


A key trend in the evolution of future \gls{6g} networks is the integration of communication and sensing functionalities, as envisioned in \gls{isac} systems. Relying solely on communication signals such as \gls{csi} provides only a partial view of the environment, as these signals are highly sensitive to noise, blockage, and rapid mobility. To enable robust, context-aware decision-making, it is important to incorporate complementary sensing modalities such as cameras, LiDAR, or radar when available \cite{alkhateeb2023deepsense}. Visual modalities capture rich spatial and semantic information about obstacles, users, and mobility patterns that may not be inferable from wireless data alone. Recent studies have demonstrated that such multimodal fusion significantly enhances performance in wireless applications including beam management \cite{farzan_beam, ghassemi2024multi}, channel estimation \cite{farzanullah2025conditional}, localization \cite{tang2024novel}, and blockage prediction \cite{charan2022computer}. By jointly leveraging wireless and non-wireless modalities, future \gls{ai}-enabled \gls{6g} networks can achieve more reliable, adaptive, and efficient operation in dynamic real-world environments.

A multimodal \gls{fm} can jointly learn from multiple data types, producing richer representations that improve performance on complex tasks \cite{fm_survey_mm}.
By combining modalities, the model can achieve more robust and context-aware inference (for example, using both wireless signal patterns and visual environmental cues for better user localization).
In the wireless domain, such models offer a scalable solution for integrating diverse sensor inputs, enabling generalization across environments and reducing reliance on extensive task-specific training data.

Motivated by the complementary strengths of multimodal data and the limited availability of labeled samples for many wireless tasks, we propose a multimodal \gls{fm} that jointly leverages wireless channel coefficients and visual input from cameras, referred to as the \gls{wmfm}. To this end, we first pretrain our model using contrastive learning \cite{chen2020simple}. 
Contrastive learning is a prominent form of self-supervised learning, 
which enables a model to learn robust representations from unlabeled data by deriving supervisory signals from the underlying data structure itself.
Specifically, contrastive learning trains models to distinguish between similar (positive) and dissimilar (negative) sample pairs, thereby learning meaningful joint representations of channels and images without requiring explicit labels. 

In the subsequent stage, we use the pretrained \gls{wmfm} as a feature extractor and attach lightweight, task-specific heads for the downstream tasks of user localization and \gls{los}/\gls{nlos} classification. The learned multimodal embeddings enable effective adaptation to diverse wireless sensing and communication tasks with minimal supervision.

The main contributions of this work are summarized as follows:
\begin{itemize}
    \item To the best of our knowledge, this work presents the first multimodal foundation model for wireless communications. The proposed \gls{wmfm} introduces a contrastive learning framework that aligns visual and wireless representations, paving the way for cross-domain feature fusion in future intelligent communication systems.
    
    \item We design modality-specific encoders for the wireless channel and image data and jointly pretrain them on a large-scale unlabeled dataset to maximize the mutual information across modalities through a contrastive objective.
    
    \item The pretrained encoders are then utilized as frozen feature extractors for downstream tasks such as localization and \gls{los}/\gls{nlos} classification, demonstrating strong generalization and sample efficiency.
    
    \item We provide a detailed theoretical analysis from both mutual information and optimization perspectives, establishing the link between contrastive pretraining and multimodal representation alignment in \gls{isac} systems.

\end{itemize}

Extensive experiments on the DeepVerse6G dataset validate the proposed framework, demonstrating that the \gls{wmfm} consistently outperforms unimodal and task-specific baselines in both \gls{los}/\gls{nlos} classification and localization tasks. 
In particular, compared to the \gls{e2e} benchmark model, the \gls{wmfm}-based approach achieves a 17\% improvement in balanced accuracy for \gls{los}/\gls{nlos} classification, while requiring approximately 90 times less training time for the lightweight task-specific head. 
Similarly, for localization, the \gls{wmfm}-based model achieves a 48.5\% reduction in Euclidean distance error compared to the \gls{e2e} benchmark, along with an 85-fold reduction in training time. 
Furthermore, the \gls{wmfm}-based heads trained on limited data even surpass the \gls{e2e} benchmark trained on the complete dataset, demonstrating the superior generalization capability and data efficiency of the proposed framework.

The rest of the paper is organized as follows: We discuss related works in Section \ref{section:relatedworks}. The system model is presented in Section \ref{section:sysmodel}, followed by our proposed model in Section \ref{section:model}. The theoretical analysis is presented in Section \ref{section:math}. Section \ref{section:dataset} discusses the dataset, and the results are discussed in Section \ref{section:results}. Finally, we conclude our paper in Section \ref{section:conclusion}.

\section{Related Work} \label{section:relatedworks}
\subsection{Foundation Models for Wireless Communications}
Inspired by the success of \glspl{fm} in \gls{nlp} and computer vision, researchers have begun exploring foundation models for wireless systems. \cite{fm_survey1} articulate the vision of a wireless physical-layer foundation model and outline the challenges in its realization, such as designing effective self-supervised tasks and handling heterogeneous time-series data. 
\cite{xu2024large} present the concept of \glspl{lmm} as universal foundation models for \gls{ai}-native wireless networks, suggesting a unified model that can understand multiple data modalities in network management and optimization contexts.
Several recent works have proposed large-scale models aimed at broad wireless tasks. For example, CommGPT \cite{jiang2025commgpt} is a graph and retrieval-augmented multimodal communication foundation model that incorporates not only wireless signal features but also side-information and knowledge graphs to enhance decision-making.

Recent efforts on wireless-specific \glspl{fm} include IQFM~\cite{mashaal2025iqfm}, a contrastive self-supervised framework that learns from raw I/Q samples for \gls{ai}-native \gls{6g} systems. IQFM uses a shared encoder with task-oriented augmentations (e.g., time shifts, channel masking) and supports fine-tuning for tasks like modulation classification, beam prediction, and RF fingerprinting. The scheme demonstrates strong few-shot generalization, showcasing the potential of large pretrained encoders in wireless applications.

The \gls{lwm}~\cite{alikhani2024large} is another wireless \gls{fm}, trained on large-scale \gls{csi} data to produce general-purpose channel embeddings. These embeddings can be repurposed for tasks like \gls{los}/\gls{nlos} classification, outperforming raw features, especially in low-data regimes. This highlights the strength of \glspl{fm} in leveraging unlabeled data for efficient transfer learning.

Recently, \glspl{fm} specifically targeting localization and sensing have been introduced. \cite{pan2025large} proposed the Large Wireless Localization Model (LWLM) for \gls{6g} positioning, which uses large-scale self-supervised learning on channel data to capture spatial radio patterns for accurate positioning. Another work, \gls{6g}~WavesFM \cite{aboulfotouh20256g} develops a unified \gls{fm} with a Vision Transformer backbone and lightweight adaptation heads (via LoRA) that is trained on real 5G data and demonstrated for joint communication and sensing tasks including 5G positioning. These efforts reflect a growing trend toward universal models that can handle RF wireless modality and tasks.

Compared to these works, our approach uniquely integrates multimodal sensor data with wireless communications data in a single \gls{fm}. While prior \glspl{fm} like IQFM and \gls{lwm} operate on RF signals alone (single-modality), we leverage the complementary information from co-existing sensor modalities, aiming to improve robustness and enable new capabilities (e.g., vision-aided communication strategies). Moreover, our work is one of the first to apply a contrastive learning framework across such heterogenous modalities (wireless channels vs. images) in the wireless domain.

\subsection{Contrastive and Multimodal Self-Supervised Learning}
Contrastive learning has been explored in the context of wireless communications and sensing. \cite{mu2025contrastive} proposed a feature-level augmentation contrastive learning scheme for wireless signal classification, demonstrating that augmenting latent features can improve representation discriminability. 
Meanwhile \cite{yancontrastive} introduced a contrastive learning framework that fuses data from multiple RF sensors for device-free indoor localization, leveraging the consistency and differences across sensors to learn more discriminative features for detecting human presence. 
These works show the versatility of contrastive learning for various wireless tasks (classification, resource allocation, sensing) and inspire our use of contrastive objectives for a multimodal wireless problem.

Extending contrastive learning to multiple modalities raises additional considerations. Recent studies have examined the interplay between single-modality and multi-modality contrastive learning.
\cite{huang2024comparison} compare multimodal and unimodal contrastive learning and show that training with multiple modalities can effectively perform a form of cross-modal canonical correlation analysis (CCA) (aligning the dominant components of each modality’s representation).
Similarly, \cite{nakada2023understanding} analyze multimodal contrastive learning and prove that, under gradient descent, optimizing a generic multimodal contrastive loss tends to align the two modality encoders such that their outputs capture the top singular vectors of the cross-covariance (i.e., the most correlated features across modalities). This suggests that multimodal contrastive training finds the common signal between modalities (analogous to what classical CCA would do), rather than just focusing on modality-specific variance.


However, to the best of our knowledge, no prior work has applied multimodal contrastive learning in the wireless domain. Existing wireless contrastive approaches remain largely unimodal without explicitly aligning different modality encoders.
These insights support our use of contrastive learning to align \gls{csi} and visual representations, ensuring the model captures shared semantic structure across heterogenous modalities.

\section{System Model}\label{section:sysmodel}

We consider an \gls{isac} scenario where each \gls{bs} is equipped with co-located sensing and communication capabilities. Specifically, each \gls{bs} has $M$ antennas, along with synchronized \gls{rgb} camera sensors. Furthermore, there are $K$ \glspl{ue} in the environment, each equipped with a single antenna.

During the uplink pilot transmission phase, each \gls{ue} transmits pilot symbols over an \gls{ofdm} frame comprising $S$ subcarriers. The received signal at the \gls{bs} for the $s^{\text{th}}$ subcarrier can be written as:
\begin{equation}
    \mathbf{y}_s = \mathbf{h}_s x_s + \mathbf{n}_s,
\end{equation}
where $x_s \in \mathbb{C}$ denotes the transmitted pilot symbol, $\mathbf{n}_s$ is additive white Gaussian noise with variance $\sigma_n^2$, and $\mathbf{h}_s \in \mathbb{C}^{M \times 1}$ represents the channel vector between the \gls{ue} and \gls{bs} antennas for the $s^{\text{th}}$ subcarrier.

For each \gls{ue}, the corresponding channel estimate across all subcarriers is given by:
\begin{equation}
    \mathbf{H} = [\,\mathbf{h}_1, \mathbf{h}_2, \ldots, \mathbf{h}_S\,] \in \mathbb{C}^{M \times S},
\end{equation}
where each row of $\mathbf{H}$ corresponds to a \gls{bs} antenna, and each column represents the complex channel coefficient\footnote{In this paper, the terms channel coefficients and \gls{csi} are used interchangeably to denote the complex-valued channel vector/matrix estimated between the \gls{ue} and \gls{bs} antennas.} at a specific subcarrier. This $\mathbf{H}$ forms the \gls{csi} modality used in the multimodal foundation model.

Simultaneously, co-located \gls{bs}-mounted \gls{rgb} cameras synchronously capture the surrounding environment with the \gls{csi} acquisition, resulting in paired sensing and communication data. Each multimodal sample can therefore be represented as:
\begin{equation}
\mathcal{D} = \{\,\mathbf{H}, \mathbf{I}_{\text{cam}}, y\,\},    
\end{equation}
where $\mathbf{I}_{\text{cam}}$ denotes the captured image, and $y$ corresponds to the downstream task label (e.g., \gls{los}/\gls{nlos} class or 3D localization). This formulation enables a unified spatial-frequency representation of the wireless environment, facilitating contrastive alignment between the radio and visual modalities during pretraining.

\subsection{Problem Formulation}
Let $\mathcal{D} = \{\mathbf{H}_i, \mathbf{I}_{\text{cam},i}, y_i\}_{i=1}^{N}$ denote the multimodal training dataset, where 
$\mathbf{H}_i \in \mathbb{C}^{M \times S}$ represents the channel coefficients corresponding to the $i^{\text{th}}$ sample,
$\mathbf{I}_{\text{cam},i}$ is the co-located \gls{rgb} image, and $y_i$ denotes the downstream label (e.g., \gls{los}/\gls{nlos} class or location).

The objective of the proposed framework is to learn modality-specific encoders 
$f_{\text{CSI}}(\cdot)$ and $f_{\text{CAM}}(\cdot)$ that project the input modalities into a common latent representation space:
\begin{equation}
    \mathbf{z}_{\text{CSI},i} = f_{\text{CSI}}(\mathbf{H}_i), \quad 
    \mathbf{z}_{\text{CAM},i} = f_{\text{CAM}}(\mathbf{I}_{\text{cam},i}),
\end{equation}
such that the learned embeddings capture both propagation-related and spatial features of the environment.
\begin{figure*}
    \centering
    \includegraphics[width=1\linewidth, trim={5mm 2mm 5mm 2mm}, clip]{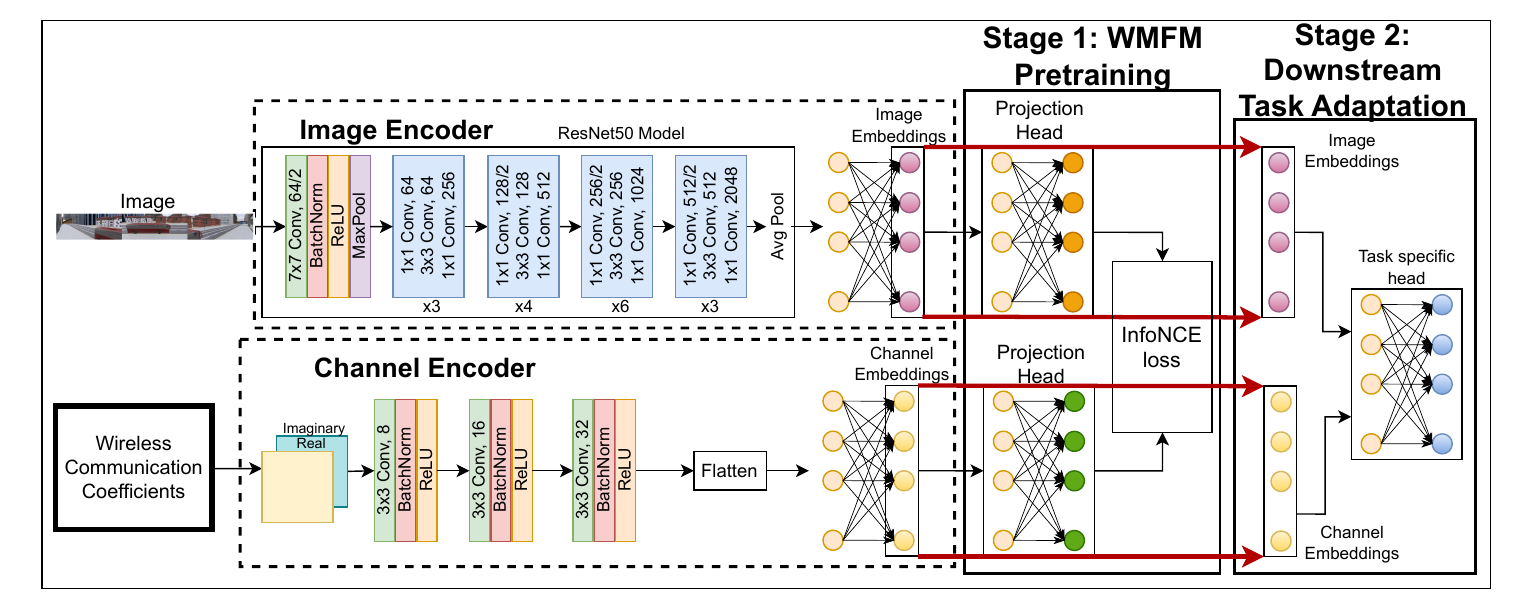}
    \caption{Architecture of our framework for \gls{wmfm} pretraining in Stage 1, and downstream task adaptation in Stage 2. The projection heads are discarded after stage 1, and the pretrained \gls{wmfm} is directly used for Stage 2.}
    \label{fig:model_architecture}
\end{figure*}

The learning objective aims to align semantically corresponding modalities e.g., \gls{csi} and camera data captured at the same time and position, while ensuring discriminability across unrelated samples. 
Formally, the training problem can be expressed as:
\begin{equation}
\begin{aligned}
    \min_{\theta_{\text{CSI}}, \theta_{\text{CAM}}} \quad & \mathcal{L}_{\text{rep}}(\mathbf{z}_{\text{CSI}}, \mathbf{z}_{\text{CAM}}) \\
    \textrm{s.t.} \quad & \|\mathbf{z}_{\text{CSI},i}\|_2 = \|\mathbf{z}_{\text{CAM},i}\|_2 = 1, \quad \forall i=1,\dots,N
\end{aligned}
\end{equation}
where $\mathcal{L}_{\text{rep}}$ denotes a representation learning loss function designed to learn a shared multimodal embedding space.
Further, $\theta_{\text{CSI}}$ and $\theta_{\text{CAM}}$ are the weights of the modality specific encoders, $f_{\text{CSI}}(\cdot)$ and $f_{\text{CAM}}(\cdot)$, respectively.
The constraints ensure that the learned embeddings are projected onto a unit hypersphere, stabilizing the training dynamics and enabling the use of cosine similarity as a distance metric.
This learned representation is later utilized for downstream tasks such as \gls{los}/\gls{nlos} classification and localization.

\section{Proposed Multimodal Foundation Model} \label{section:model}

The goal of the proposed framework is to develop a task-agnostic multimodal \gls{fm} that jointly learns representations from wireless communication and visual sensing modalities. 
Building upon the system model and problem formulation described earlier, the model is designed to extract complementary information from both the radio and visual domains, thereby enabling a unified latent representation that generalizes across diverse downstream \gls{isac} tasks such as \gls{los}/\gls{nlos} classification and localization.

To achieve this, the framework follows a two-stage training paradigm. 
In the first stage, self-supervised learning pretraining is performed to align modality-specific features in a shared embedding space using large-scale unlabeled multimodal data. 
The overall pipeline consists of two parallel encoders. One for wireless channel coefficients and one for \gls{rgb} camera images, each followed by a lightweight projection head, as illustrated in Fig.~\ref{fig:model_architecture}. 
In the second stage, the pretrained encoders are frozen and used as feature extractors, while lightweight task-specific heads are trained using task-specific labels.
The following subsections describe the architectural components, and pretraining methodology, in detail.

\subsection{Model Architecture Overview}
The proposed multimodal foundation model consists of two modality-specific encoders that extract high-level representations from wireless and visual domains, respectively, followed by projection heads that map the modality embeddings into a common latent space, as illustrated in Fig.~\ref{fig:model_architecture}.

The image encoder employs a ResNet50 backbone that processes the \gls{rgb} frame captured by the \gls{bs}-mounted camera. The convolutional layers progressively reduce the spatial resolution while increasing the feature dimensionality. The output feature map is averaged through a global average pooling layer, followed by a single \gls{mlp} layer to generate the image embedding vector $\mathbf{z}_{\text{CAM}} \in \mathbb{R}^{d}$.

The channel encoder processes the complex-valued wireless channel coefficients by first separating the real and imaginary components into two real-valued matrices. These matrices are then concatenated along the depth dimension to form a dual-channel input tensor for the \gls{cnn}. This results in a real-valued input tensor of size \( \mathbf{H} \in \mathbb{R}^{M \times S \times 2} \).
It is then passed through a series of $3 \times 3$ convolutional blocks with batch normalization and ReLU activation to extract local spatial-frequency features. The final feature map is flattened and linearly projected to obtain the channel embedding $\mathbf{z}_{\text{CSI}} \in \mathbb{R}^{d}$.

Both encoders are followed by lightweight projection heads composed of two fully connected layers with ReLU activation. These heads map the modality-specific embeddings into a shared latent space where the contrastive objective is applied. The use of projection heads serves multiple purposes. First, it decouples the representation learning space from the contrastive optimization space, allowing the encoders to focus on extracting general, transferable features rather than overfitting to the contrastive task \cite{huang2024comparison}. 
Secondly, applying the contrastive loss on the projected embeddings rather than directly on the encoder outputs has been shown to improve both stability during training and downstream task generalization \cite{huang2024comparison}. After pretraining, the projection heads are discarded, and the encoder embeddings are utilized for downstream adaptation.

\subsection{Self-Supervised Pretraining with Contrastive Learning}
During pretraining, the objective is to align representations of temporally and spatially corresponding modalities (e.g., \gls{csi} and camera data captured at the same time and \gls{bs}) while pushing apart embeddings from unrelated samples. Given a batch of $N$ paired samples $\{(\mathbf{H}_i, \mathbf{I}_{\text{cam},i})\}_{i=1}^{N}$, the embeddings produced by the two encoders are denoted as $\mathbf{z}_{\text{CSI},i}$ and $\mathbf{z}_{\text{CAM},i}$, respectively. 

A contrastive loss based on the symmetric \gls{infonce} formulation is applied \cite{oord2018representation}:
{\small
\begin{align}
\mathcal{L}_{\text{NCE}}^{\text{sym}} =
\frac{1}{2N}\sum_{i=1}^{N}\Bigg[
& -\log \frac{\exp\left(\mathrm{sim}(\mathbf{z}_{\mathrm{CSI},i},\mathbf{z}_{\mathrm{CAM},i})/\tau\right)}{\sum_{j=1}^{N}\exp\left(\mathrm{sim}(\mathbf{z}_{\mathrm{CSI},i},\mathbf{z}_{\mathrm{CAM},j})/\tau\right)} \notag \\
& -\log \frac{\exp\left(\mathrm{sim}(\mathbf{z}_{\mathrm{CAM},i},\mathbf{z}_{\mathrm{CSI},i})/\tau\right)}{\sum_{j=1}^{N}\exp\left(\mathrm{sim}(\mathbf{z}_{\mathrm{CAM},i},\mathbf{z}_{\mathrm{CSI},j})/\tau\right)}
\Bigg]
\label{eq:symmetric_infonce}
\end{align}
}

where $\mathrm{sim}(\cdot,\cdot)$ denotes the similarity between embeddings, and $\tau$ is the temperature parameter that controls the sharpness of the similarity distribution.

The symmetric form of the \gls{infonce} loss jointly optimizes both cross-modal directions: \gls{csi}$\rightarrow$CAM and CAM$\rightarrow$\gls{csi}. The first term treats \gls{csi} embeddings as anchors and camera embeddings as targets, while the second term reverses the roles. The bidirectional loss formulation ensures that both modality encoders contribute equally to the optimization, preventing feature dominance from a single modality and promoting a well-balanced joint embedding space \cite{radford2021learning}. By contrasting positive pairs against all other negatives in the batch, the model learns to encode both spatial and propagation characteristics in a unified representation space. 
The similarity metric is defined as the cosine similarity between two $\ell_2$-normalized embeddings \cite{oord2018representation}: 
\begin{equation}
\mathrm{sim}(\mathbf{z}_a, \mathbf{z}_b)
= \frac{\mathbf{z}_a^{\top}\mathbf{z}_b}{\|\mathbf{z}_a\|_2\,\|\mathbf{z}_b\|_2}.
\label{eq:similarity_metric}
\end{equation}

\begin{figure}
    \centering
    \includegraphics[width=1\linewidth, trim={5mm 2mm 5mm 2mm}, clip]{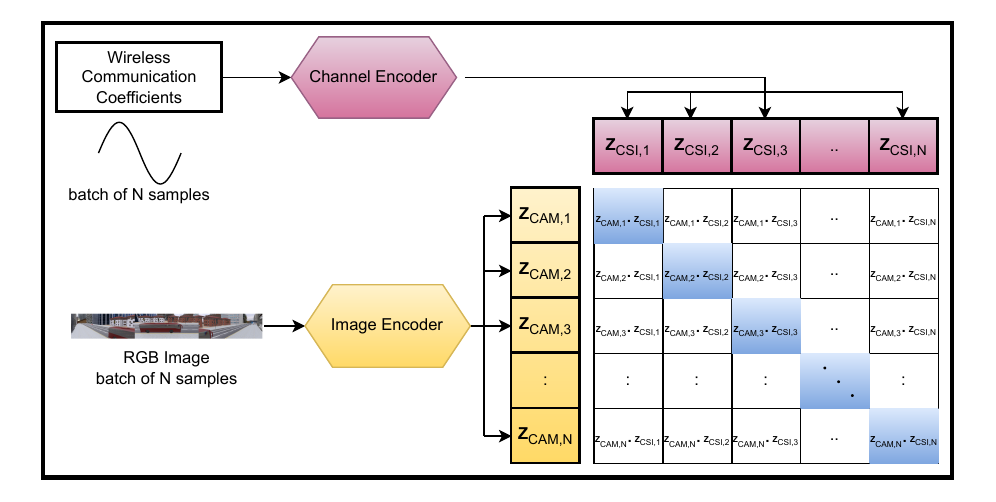}
    \caption{Dot-product similarity matrix enabling contrastive alignment of wireless channel and image embeddings.}
    \label{fig:contrastive}
\end{figure}
Fig.~\ref{fig:contrastive} illustrates the cross-modal similarity computation between the wireless channel embeddings and the corresponding camera-image embeddings. 
Given a batch of $N$ samples, the channel encoder maps each complex channel coefficient tensor into a latent representation $\mathbf{z}_{\text{CSI},i}$, while the image encoder extracts visual features $\mathbf{z}_{\text{CAM},j}$ from the associated scene. 
A pairwise dot product is then computed between every $\mathbf{z}_{\text{CSI},i}$ and $\mathbf{z}_{\text{CAM},j}$, forming an $N \times N$ similarity matrix. 
Diagonal elements (highlighted in blue) correspond to positive pairs originating from the same physical scene (i.e., captured at the same \gls{bs} and time), whereas off-diagonal entries represent negative pairs that enforce discriminative alignment during contrastive learning.
In essence, contrastive learning pulls positive pairs closer in the shared embedding space while simultaneously pushing apart negative pairs, thereby shaping a representation that preserves cross-modal semantic consistency.

The pretraining phase is fully self-supervised and does not rely on downstream labels, enabling large-scale multimodal representation learning using unlabeled \gls{csi}-camera pairs.

\section{Mathematical and Theoretical Analysis}\label{section:math}
This section provides a formal analysis of the proposed learning objective, establishing its connection to mutual information maximization and cross-modal representation alignment.
It should be noted that the fundamental mathematical derivations in Sections \ref{Section:Math:MI} and \ref{Section:Math:Opt} are based on the work of  \cite{oord2018representation} and \cite{wang2020understanding}. Our contribution lies in extending this analysis to the specific context of \gls{wmfm}, verifying that these bounds hold for the joint distribution of channel coefficients and visual imagery.

\subsection{Mutual Information Perspective} \label{Section:Math:MI}

In the proposed \gls{wmfm} pretraining, the contrastive \gls{infonce} loss can be interpreted as a surrogate for maximizing the mutual information 
between the \gls{csi} and image modalities. The mutual information $I(X;Y)$ quantifies how much knowing one variable reduces uncertainty about the other. 
Formally, it is defined as:
\begin{equation}
    I(X;Y) = \mathbb{E}_{p(x,y)}\left[\log \frac{p(x,y)}{p(x)\,p(y)}\right].
\end{equation}
We aim to maximize $I(Z_{\mathrm{CSI}};\,Z_{\mathrm{CAM}})
$ 
to enforce cross-modal representation alignment. Inspired from \cite{oord2018representation}, we aim to derive a lower bound on the mutual information between wireless and visual modalities, in terms of our loss function.

The \gls{infonce} loss, computed by anchoring the channel input \(\textbf{H} \) and contrasting it against image samples \( \textbf{I} \), is given by
\begin{equation}
\mathcal{L}_{\textbf{H} \rightarrow \textbf{I}} = -\mathbb{E}_i \left[ \log \frac{s(\mathbf{z}_{\text{CSI},i}, \mathbf{z}_{\text{CAM},i})}{\sum_{k=1}^{N} s(\mathbf{z}_{\text{CSI},i}, \mathbf{z}_{\text{CAM},k})} \right] \label{Eq:MI:infonce}
\end{equation}
where \( s(\mathbf{z}_{\text{CSI},i}, \mathbf{z}_{\text{CAM},i}) \) denotes the similarity score between the \gls{csi} embedding of \( \textbf{H}_i \) and the image embedding of \( \textbf{I}_i \).
The scoring function $s(\cdot, \cdot)$ measures the similarity or compatibility between two modality embeddings, typically via cosine similarity.
Optimizing the \gls{infonce} loss in the proposed framework can be interpreted as learning to classify a positive image embedding \( \textbf{z}_i^{\mathrm{CAM}} \) that corresponds to a given \gls{csi} embedding \( \textbf{z}_i^{\mathrm{CSI}} \). Let \( d = i \) denote the event that the sample \( i \) is the correct (positive) match. Then, the probability of selecting the correct pair from a batch of size \( N \) is given by:
\begin{equation} \label{Eq:MI:prob1}
    p(d = i \mid \mathcal{Z}_{\mathrm{CAM}}, \textbf{z}_i^{\mathrm{CSI}}) 
    = \frac{p(\textbf{z}_i^{\mathrm{CAM}} \mid \textbf{z}_i^{\mathrm{CSI}}) \prod_{j \ne i} p(\textbf{z}_j^{\mathrm{CAM}})}{\sum\limits_{j=1}^{N} p(\textbf{z}_j^{\mathrm{CAM}} \mid \textbf{z}_i^{\mathrm{CSI}}) \prod_{k \ne j} p(\textbf{z}_k^{\mathrm{CAM}})}.
\end{equation}




The Bayes' rule can be used to simplify the Eq.  \eqref{Eq:MI:prob1} to:
\begin{equation} \label{Eq:MI:Bayes}
    p(d = i \mid \mathcal{Z}_{\mathrm{CAM}}, \textbf{z}_i^{\mathrm{CSI}}) 
    = \frac{ \frac{p(\textbf{z}_i^{\mathrm{CAM}} \mid \textbf{z}_i^{\mathrm{CSI}})}{p(\textbf{z}_i^{\mathrm{CAM}})} }{ \sum\limits_{j=1}^{N} \frac{p(\textbf{z}_j^{\mathrm{CAM}} \mid \textbf{z}_i^{\mathrm{CSI}})}{p(\textbf{z}_j^{\mathrm{CAM}})} }.
\end{equation}

Therefore, the optimal scoring function \( s^*(\textbf{z}_i^{\mathrm{CSI}}, \textbf{z}_j^{\mathrm{CAM}}) \) that maximizes the probability of selecting the correct match is proportional ($\propto$) to the true likelihood ratio:
\begin{equation} \label{Eq:MI:prop}
    s^*(\textbf{z}_i^{\mathrm{CSI}}, \textbf{z}_j^{\mathrm{CAM}}) \propto \frac{p(\textbf{z}_j^{\mathrm{CAM}} \mid \textbf{z}_i^{\mathrm{CSI}})}{p(\textbf{z}_j^{\mathrm{CAM}})}.
\end{equation}

This confirms that the \gls{infonce} loss implicitly estimates a density ratio between the joint and marginal distributions over embeddings, and encourages the encoders to learn mutually informative representations.

Inserting Eq. \eqref{Eq:MI:prop} into the \gls{infonce} loss Eq. \eqref{Eq:MI:infonce}, the optimal loss becomes:

\begin{equation}
\resizebox{\columnwidth}{!}{$
\begin{aligned}
\mathcal{L}_{\textbf{H}\rightarrow \textbf{I}} &= -\mathbb{E}_{\mathcal{Z}_{\mathrm{CAM}}} \left[ \log \frac{ \frac{p(\textbf{z}_{\mathrm{CAM}}|\textbf{z}_{\mathrm{CSI}})}{p(\textbf{z}_{\mathrm{CAM}})} }{ \frac{p(\textbf{z}_{\mathrm{CAM}}|\textbf{z}_{\mathrm{CSI}})}{p(\textbf{z}_{\mathrm{CAM}})} + \sum\limits_{j \in \mathcal{X}_{\text{neg}}} \frac{p(\textbf{z}_j|\textbf{z}_{\mathrm{CSI}})}{p(\textbf{z}_j)} } \right] \\
&= \mathbb{E}_{\mathcal{Z}_{\mathrm{CAM}}} \left[ \log \left( 1 + \frac{p(\textbf{z}_{\mathrm{CAM}})}{p(\textbf{z}_{\mathrm{CAM}}|\textbf{z}_{\mathrm{CSI}})} \sum_{j=1}^{N-1} \frac{p(\textbf{z}_j|\textbf{z}_{\mathrm{CSI}})}{p(\textbf{z}_j)} \right) \right] \\
&\approx \mathbb{E}_{\mathcal{Z}_{\mathrm{CAM}}} \left[ \log \left( 1 + \frac{p(\textbf{z}_{\mathrm{CAM}})}{p(\textbf{z}_{\mathrm{CAM}}|\textbf{z}_{\mathrm{CSI}})} (N - 1) \mathbb{E}_{\textbf{z}_j} \left[ \frac{p(\textbf{z}_j|\textbf{z}_{\mathrm{CSI}})}{p(\textbf{z}_j)} \right] \right) \right] \\
&= \mathbb{E}_{\mathcal{Z}_{\mathrm{CAM}}} \left[ \log \left( 1 + \frac{p(\textbf{z}_{\mathrm{CAM}})}{p(\textbf{z}_{\mathrm{CAM}}|\textbf{z}_{\mathrm{CSI}})} (N - 1) \right) \right] \\
&\ge \mathbb{E}_{\mathcal{Z}_{\mathrm{CAM}}} \left[ \log \left( \frac{p(\textbf{z}_{\mathrm{CAM}})}{p(\textbf{z}_{\mathrm{CAM}}|\textbf{z}_{\mathrm{CSI}})} N \right) \right] \\
&= -I(Z_{\mathrm{CAM}}, Z_{\mathrm{CSI}}) + \log N.
\end{aligned}
$}
\end{equation}

Therefore, we obtain a lower bound on mutual information:
\begin{equation}
    I(Z_{\mathrm{CAM}}, Z_{\mathrm{CSI}}) \ge \log(N) - \mathcal{L}_{\textbf{H}\rightarrow \textbf{I}}. \label{eq:infonce_mi_bound}
\end{equation}

This bound implies that increasing the number of negative samples \( N \) tightens the estimate of mutual information. Simultaneously, minimizing \( \mathcal{L}_{\textbf{H}\rightarrow \textbf{I}} \) encourages the encoder networks to produce aligned and informative representations across the wireless and visual modalities.
Our loss is implemented in a symmetric form~\cite{radford2021learning}, optimizing the contrastive objective in both directions: from \gls{csi} to image and from image to \gls{csi}.

\subsection{Optimization Perspective: Alignment and Uniformity}\label{Section:Math:Opt}
Beyond the mutual information view, the gradient dynamics of contrastive objectives reveal how they induce desirable representation properties, notably alignment of positive pairs and uniformity of embeddings on the unit sphere.
Following \cite{wang2020understanding}, we characterize these properties as follows: alignment minimizes the expected distance between positive pairs ($\mathbb{E}[\|\mathbf{z}_{\text{CSI}} - \mathbf{z}_{\text{CAM}}\|_2^2]$), while uniformity encourages the embeddings to maximize the entropy of the induced distribution on the hypersphere.

Considering the \gls{csi} embedding as the anchor $\textbf{z}_{\text{CSI},i}$ and its corresponding camera embedding $\textbf{z}_{\text{CAM},i}$ as the positive sample, we define:
\[
p_i = \frac{\exp(\mathrm{sim}(\textbf{z}_{\text{CSI},i},\textbf{z}_{\text{CAM},i})/\tau)}{\sum_{j=1}^{N}\exp(\mathrm{sim}(\textbf{z}_{\text{CSI},i},\textbf{z}_{\text{CAM},j})/\tau)},
\]
where $p_i$ denotes the softmax probability of the true positive pair and $p_j$ corresponds to a negative pair $j\neq i$. Differentiating the \gls{infonce} loss equation yields
\begin{align}
\frac{\partial L_{\text{NCE}}}{\partial\,\mathrm{sim}(\textbf{z}_{\text{CSI},i}, \textbf{z}_{\text{CAM},i})} &= -\frac{1 - p_i}{\tau} < 0, \label{eq:grad-pos-csi}\\
\frac{\partial L_{\text{NCE}}}{\partial\,\mathrm{sim}(\textbf{z}_{\text{CSI},i}, \textbf{z}_{\text{CAM},j})} &= \frac{p_j}{\tau} > 0,\quad j\neq i. \label{eq:grad-neg-csi}
\end{align}
These gradients characterize the optimization forces acting on embeddings: for the positive pair, the derivative is negative, indicating that increasing the similarity $\mathrm{sim}(\textbf{z}_{\text{CSI},i},\textbf{z}_{\text{CAM},i})$ reduces the loss, thereby pulling the \gls{csi} and camera embeddings closer (alignment). For each negative pair, the derivative is positive, meaning that increasing their similarity increases the loss; gradient descent thus pushes them apart (uniformity). The strength of each repulsive term scales with $p_j$, such that negatives most similar to the anchor (i.e., hard negatives) are pushed away more strongly, balancing attraction and repulsion dynamically during training.

This implicit trade-off ensures that \gls{csi} and camera embeddings corresponding to the same scene are pulled together, while embeddings from different samples are evenly dispersed on the hypersphere. Following \cite{wang2020understanding}, at the optimum, positive pairs collapse to identical points (perfect alignment), and embeddings of distinct instances become maximally separated, approaching orthogonality as the embedding dimension increases. These coupled forces yield well-disentangled and informative multimodal representations in $\mathcal{Z}$, forming a robust foundation for downstream \gls{isac} adaptation tasks.

\subsection{Theoretical Guarantee for WMFM Pretraining}

To ground our framework in formal guarantees, we adapt the results of \cite{saunshi2019} to our multimodal contrastive learning setup. The following theorem illustrates how low InfoNCE loss ensures downstream linear predictability in the learned embedding space.

\begin{theorem}[Generalization of WMFM Embeddings via InfoNCE]
Let $(\textbf{H}, \textbf{I}_{\mathrm{cam}}, y) \sim \mathcal{D}$ be paired data sampled from a multimodal distribution, where each pair is collected simultaneously from the same base station and thus shares a common latent scene class \( y \). Let \( \textbf{z}_{\mathrm{CSI}} = f_{\mathrm{CSI}}(\textbf{H}) \) and \( \textbf{z}_{\mathrm{CAM}} = f_{\mathrm{CAM}}(\textbf{I}_{\mathrm{cam}}) \) be the $\ell_2$-normalized embeddings produced by the \gls{wmfm} encoders, trained using the symmetric \gls{infonce} loss over \( N \) negatives per batch.

Assume the encoders \( f_{\mathrm{CSI}} \) and \( f_{\mathrm{CAM}} \) are \( L \)-Lipschitz, and that the training loss satisfies \( \mathcal{L}_{\mathrm{InfoNCE}} \le \epsilon \). Then, with high probability over training data, there exists a linear classifier \( g \) such that the classification error of predicting \( y \) from either embedding is bounded by:
\[
\mathbb{P}_{(\textbf{H}, \textbf{I}, y)}\left[ g(\textbf{z}_{\mathrm{CSI/CAM}}) \ne y \right] \le \mathcal{O}(\sqrt{\epsilon} + \delta),
\]
where \( \delta \) captures inter-class overlap and the impact of modality-specific noise.
\end{theorem}

The proof follows from \cite{saunshi2019}, which shows that low \gls{infonce} loss leads to linearly separable embeddings. Since each CSI–image pair corresponds to the same scene class by construction, and the encoders are Lipschitz, small input noise (e.g., in \gls{csi}) yields bounded changes in embeddings. The contrastive loss promotes intra-class similarity and inter-class separation, enabling a linear classifier, such as \gls{mlp} to recover the latent classes.

\subsection{Complexity Analysis}

\label{sec:complexity-analysis}

We analyze the computational cost of the \gls{wmfm} contrastive pretraining pipeline by examining both encoder complexity and \gls{infonce} loss computation. The image encoder used is ResNet-50, containing approximately 23.5 million parameters, while the \gls{csi} encoder is significantly lighter, with around 4.7 million parameters. This encoder asymmetry enables efficient processing of wireless data without compromising visual feature richness.

During contrastive training, pairwise similarities must be computed across all \(N\) samples in a batch. This results in a per-step complexity of \(\mathcal{O}(N^2 d)\) operations to compute the similarity matrix, where \(d\) is the embedding dimension. The associated memory requirement is also \(\mathcal{O}(N^2)\), since the full similarity matrix must be stored. 

Overall, the total computational cost per training step is \(\mathcal{O}(N \cdot C_{\text{enc}} + N^2 d)\), where \(C_{\text{enc}}\) denotes the average encoder cost per sample. For practical batch sizes (e.g., \(N \leq 512\)), this cost remains tractable on modern GPUs. At inference time, only a single forward pass through the \gls{csi} and image encoder is required, enabling low-latency, real-time deployment.

\section{Dataset and Scenario} \label{section:dataset}
\subsection{DeepVerse6G Multimodal Wireless Dataset}

\begin{figure}
    \centering
    \includegraphics[width=1\linewidth]{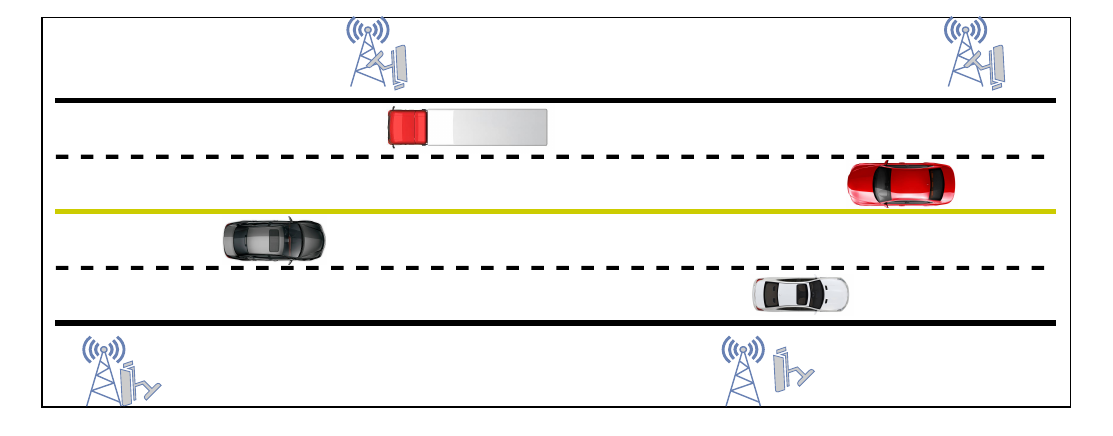}
    \caption{The O1 scenario within the DeepVerse dataset \cite{DeepVerse} features a highway environment served by four \gls{bs}. Each \gls{bs} is equipped with three cameras to ensure comprehensive visual coverage of the road segment.}
    \label{fig:O1Scenario}
\end{figure}

In this work, we employ the publicly available DeepVerse6G \cite{DeepVerse} dataset to evaluate the performance of the proposed model. 
DeepVerse6G is a large-scale simulated multimodal dataset that leverages ray-tracing techniques to generate physically realistic wireless propagation environments and corresponding multimodal sensor data. 
For the experimental analysis, we focus on the O1 scenario, which represents an urban street environment consisting of four lanes (two in each direction). 
Along this street, four \glspl{bs} are deployed, each equipped with multiple sensing modalities including cameras, LiDAR, and radar. 
Each \gls{bs} is equipped with three \gls{rgb} cameras oriented to cover the left, center, and right sides of the street, thereby ensuring full visual coverage of the vehicular environment. 
In this study, we restrict our analysis to the wireless communication data and camera images, which serve as the primary modalities for self-supervised pretraining and downstream evaluation.

\subsection{\gls{isac} Scenario and Downstream Tasks}

We evaluate the proposed framework on two downstream \gls{isac} tasks that exploit the complementarity between radio propagation cues (\gls{csi}) and visual context (camera). 
In all experiments, the pretrained encoders are frozen and used as feature extractors, and only lightweight task heads are trained. 
We concatenate the modality embeddings $\mathbf{z}_{\mathrm{CSI}}$ and $\mathbf{z}_{\mathrm{CAM}}$ as $\mathbf{z}=[\mathbf{z}_{\mathrm{CAM}};\mathbf{z}_{\mathrm{CSI}}]\in\mathbb{R}^{2d}$ and feed $\mathbf{z}$ to task-specific adapters that include modality-wise feature extractors and a simple two-weight attention to adaptively fuse image and channel features.

\subsubsection{Task 1: \gls{los}/\gls{nlos} Classification}
\gls{los}/\gls{nlos} classification plays a crucial role in modern wireless systems, as it directly impacts key operations such as localization, beamforming, and channel estimation. Accurately distinguishing between these two propagation conditions enables the design of robust communication strategies that can mitigate multipath fading, reduce latency, and enhance throughput \cite{alikhani2024large}. In emerging \gls{6g} networks, where high-frequency signals are more susceptible to blockage and reflection, reliable \gls{los}/\gls{nlos} identification becomes even more critical for maintaining quality of service and ensuring seamless connectivity.

The \gls{los} class constitutes the majority (93.5\%) in our highway-like O1 scenario, which induces severe class imbalance. 
To mitigate the dominance of easy \gls{los} samples and focus learning on informative \gls{nlos} cases, we adopt the focal loss \cite{focal_loss}. 
Concretely, the task head consists of two separate \glspl{mlp} that first refine the image and channel halves of the joint embedding $\mathbf{z}$. The resulting features are then concatenated and processed through a lightweight transformer module, which models inter-modal interactions and contextual dependencies. The transformer's output is subsequently passed to a final \gls{mlp} classifier that produces the logits $\boldsymbol{\ell} \in \mathbb{R}^2$.

Let $y_i\in\{0,1\}$ be the ground-truth label (\gls{nlos}/\gls{los}) and $\mathbf{p}_i=\mathrm{softmax}(\boldsymbol{\ell}_i)$ the predicted class probabilities. 
The multi-class focal loss (applied here with 2 classes) is:
\begin{equation}
\mathcal{L}_{\text{focal}}
= -\frac{1}{N}\sum_{i=1}^{N}\sum_{c=1}^{2}
\alpha_c\,(1-p_{i,c})^{\gamma}\,\mathbf{1}[y_i{=}c]\;\log p_{i,c},
\label{eq:focal}
\end{equation}
where $p_{i,c}$ is the predicted probability for class $c$, $\gamma{>}0$ down-weights easy examples, and $\alpha_c$ balances class priors (higher for the minority \gls{nlos} class). 

We minimize the focal loss $\mathcal{L}_{\text{focal}}$ to train only the classifier and attention parameters while keeping the encoders fixed:
\begin{equation}
\min_{\;\theta_{\text{head}}}\;\;\mathcal{L}_{\text{focal}}(\theta_{\text{head}};\,\mathbf{z}_{\mathrm{CAM}},\mathbf{z}_{\mathrm{CSI}},y).
\end{equation}
This task reflects the \emph{communication} facet of \gls{isac}, where propagation condition recognition (\gls{los}/\gls{nlos}) aids link adaptation.

\subsubsection{Task 2: Localization/Positioning}
Accurate localization is a fundamental requirement for many wireless applications, including autonomous navigation, intelligent transportation systems, and context-aware services. In \gls{6g} and beyond, precise user and device positioning enables highly directional beamforming, efficient resource allocation, and seamless handover between cells \cite{localization}. 
Therefore, designing robust localization systems that can operate reliably in diverse channel conditions is essential for the next generation of intelligent wireless networks.

We estimate the 3D user position with a hybrid objective: continuous regression for the longitudinal coordinate $x$, and discrete classification for lateral lane index $y$ (4 classes) and height/type $z$ (9 classes). 
Empirically, discretizing $y$ and $z$ improves robustness and interpretability. 
The task head follows the hybrid design: modality-wise \glspl{mlp} $\!\rightarrow$ attention fusion $\!\rightarrow$ shared features $\!\rightarrow$ three heads producing $x$-regression and $(y,z)$ logits.
The total localization loss $\mathcal{L}_{\text{loc}}$ is a weighted sum of the losses of the 3 coordinates.
\begin{equation}
\mathcal{L}_{\text{loc}} = \lambda_x\,\mathcal{L}_{x} + \lambda_y\,\mathcal{L}_{y} + \lambda_z\,\mathcal{L}_{z},
\quad \lambda_x,\lambda_y,\lambda_z \ge 0,
\label{eq:loc_total}
\end{equation}
where $\mathcal{L}_{x}$, $\mathcal{L}_{y}$, and $\mathcal{L}_{z}$ are the losses for $x$, $y$, and $z$ coordinate, respectively.
We optimize
\begin{equation}
\min_{\;\theta_{\text{head}}}\;\;\mathcal{L}_{\text{loc}}(\theta_{\text{head}};\,\mathbf{z}_{\mathrm{CAM}},\mathbf{z}_{\mathrm{CSI}}, x,y,z).
\end{equation}
This task emphasizes the sensing facet of \gls{isac} by leveraging radio–visual fusion to infer user geometry (longitudinal position and lane/height classes).

In both tasks, the transfomer attention gate provides a simple fusion mechanism that adapts the contribution of visual context (scene layout, occlusions) and radio cues (multipath structure, \gls{los}/\gls{nlos}) to the downstream objective. 
The frozen, contrastively pretrained encoders supply a modality-invariant representation, while the task heads specialize this representation to communication(\gls{los}/\gls{nlos}) and sensing (localization) objectives with minimal supervision.

\section{Experimental Results} \label{section:results}

The O1 scenario of the DeepVerse6G dataset was partitioned into training, validation, and testing sets in a 33\%–33\%–33\% split. Each \gls{bs} was equipped with $M = 16$ antennas, while each \gls{ue} is equipped with a single antenna. The number of subcarriers was set to $K = 8$. The career frequency was set as 28 GHz.
Each \gls{bs} has three cameras (left, center, right). Their images are horizontally concatenated and downsampled by 4× in both dimensions to reduce computation while retaining key visual details.

The proposed \gls{wmfm} model was pretrained using a contrastive learning framework based on the \gls{infonce} loss. The image encoder utilized a ResNet-50 backbone, which was fine-tuned during the pretraining phase. ResNet-50 is a deep \gls{cnn} composed of an initial convolution and max-pooling layer, followed by 16 residual blocks organized into four stages, and terminating with global average pooling. A two-layer \gls{mlp} with 1024 and $d=128$ neurons was appended to generate the image embeddings from the ResNet-50 features.
For the \gls{csi} modality, complex-valued channel coefficients were separated into real and imaginary parts and processed through a series of \gls{cnn} blocks, as illustrated in Fig.~\ref{fig:model_architecture}. The output feature maps were flattened and passed through a two-layer \gls{mlp} with 32 and $d=128$ neurons. Both image and channel encoders were followed by lightweight projection heads, each consisting of two-layer \glspl{mlp} with 256 and 32 neurons. The temperature parameter used in the symmetric \gls{infonce} loss in Eq.~\eqref{eq:symmetric_infonce} was empirically set to $\tau = 0.1$, and the batch size $N$ was set to 64.

The pretraining of \gls{wmfm} was conducted for 60 epochs. The initial learning rate was set to $1 \times 10^{-3}$ and was reduced by a factor of 10 if the validation loss did not improve for five consecutive epochs. The Adam optimizer was employed for model optimization. The training took around 7 hours on a A100 GPU.

As a baseline, we implement an \gls{e2e} benchmark model where identical channel and image encoders, along with the task-specific head, are trained from scratch using only supervised data. Unlike the proposed \gls{wmfm} which leverages pretrained modality encoders, the \gls{e2e} model learns all parameters jointly without any pretraining, serving as a reference to evaluate the benefits of multimodal representation learning.

\subsection{Training Results}

\begin{figure}
    \centering
    \includegraphics[width=\linewidth]{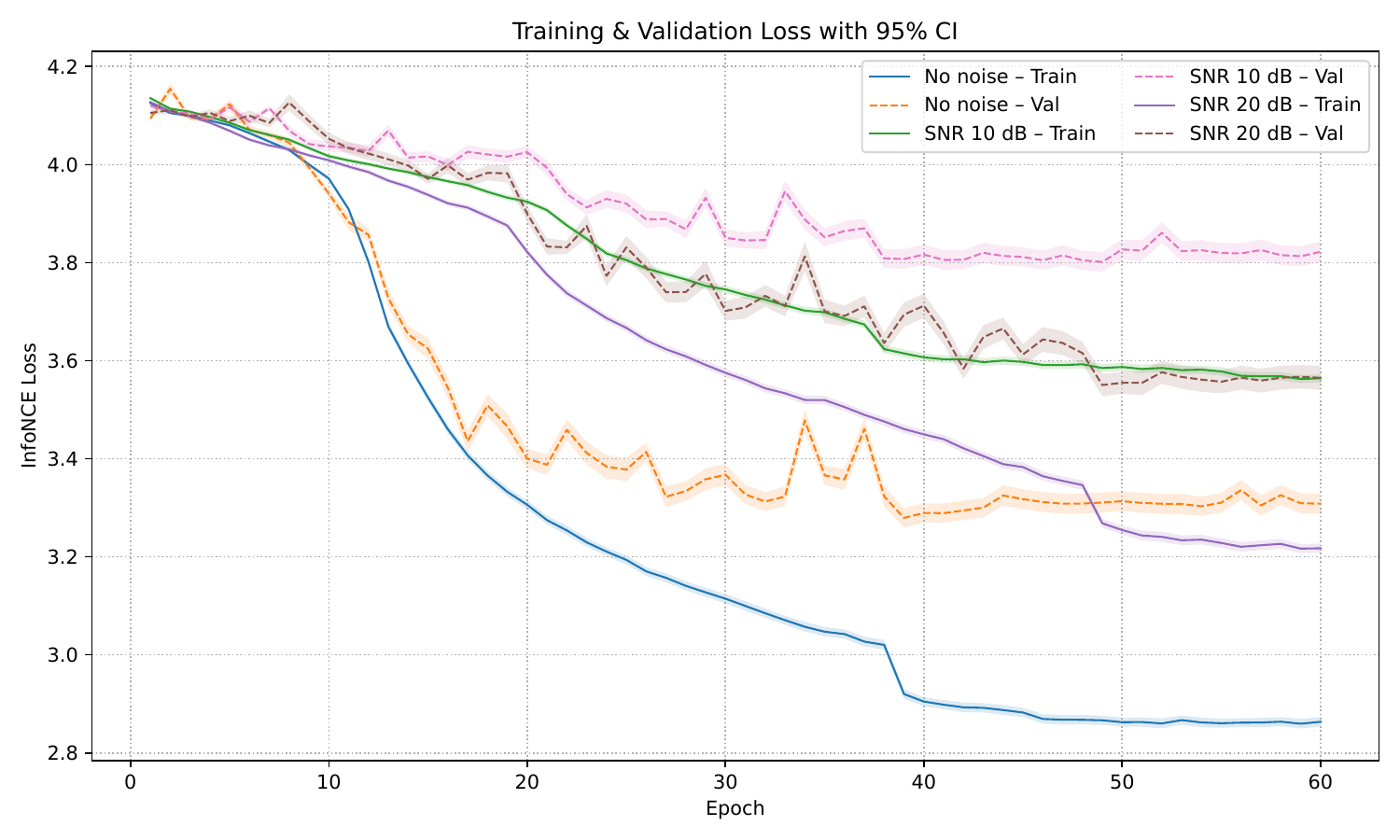}
    \caption{Training and validation \gls{infonce} loss for different noise levels.
    }
    \label{fig:training:epoch}
\end{figure}
Fig. ~\ref{fig:training:epoch} illustrates the evolution of the training and validation losses during the contrastive pretraining phase using the \gls{infonce} objective across different noise levels applied to the \gls{csi} modality. In particular, \SI{20}{\dB} and \SI{10}{\dB} refer to noise added such that the resulting \gls{snr} had a mean of \SI{20}{\dB} and \SI{10}{\dB}, respectively, both with a standard deviation of \SI{10}{\dB}.
The no-noise model shows faster and more stable convergence, indicating effective cross-modal representation learning under clean conditions. 
Introducing additive noise results in a slower convergence and slightly higher loss values; however, as shown in subsequent results, noisy pretraining improves the model’s robustness and generalization under low-\gls{snr} conditions. 
Apart from the analysis presented in Fig.~\ref{fig:loc:noise}, the no-noise pretrained model is used for all subsequent evaluations.




\begin{figure}
    \centering

    \begin{subfigure}{\columnwidth}
        \centering
        \includegraphics[width=\linewidth]{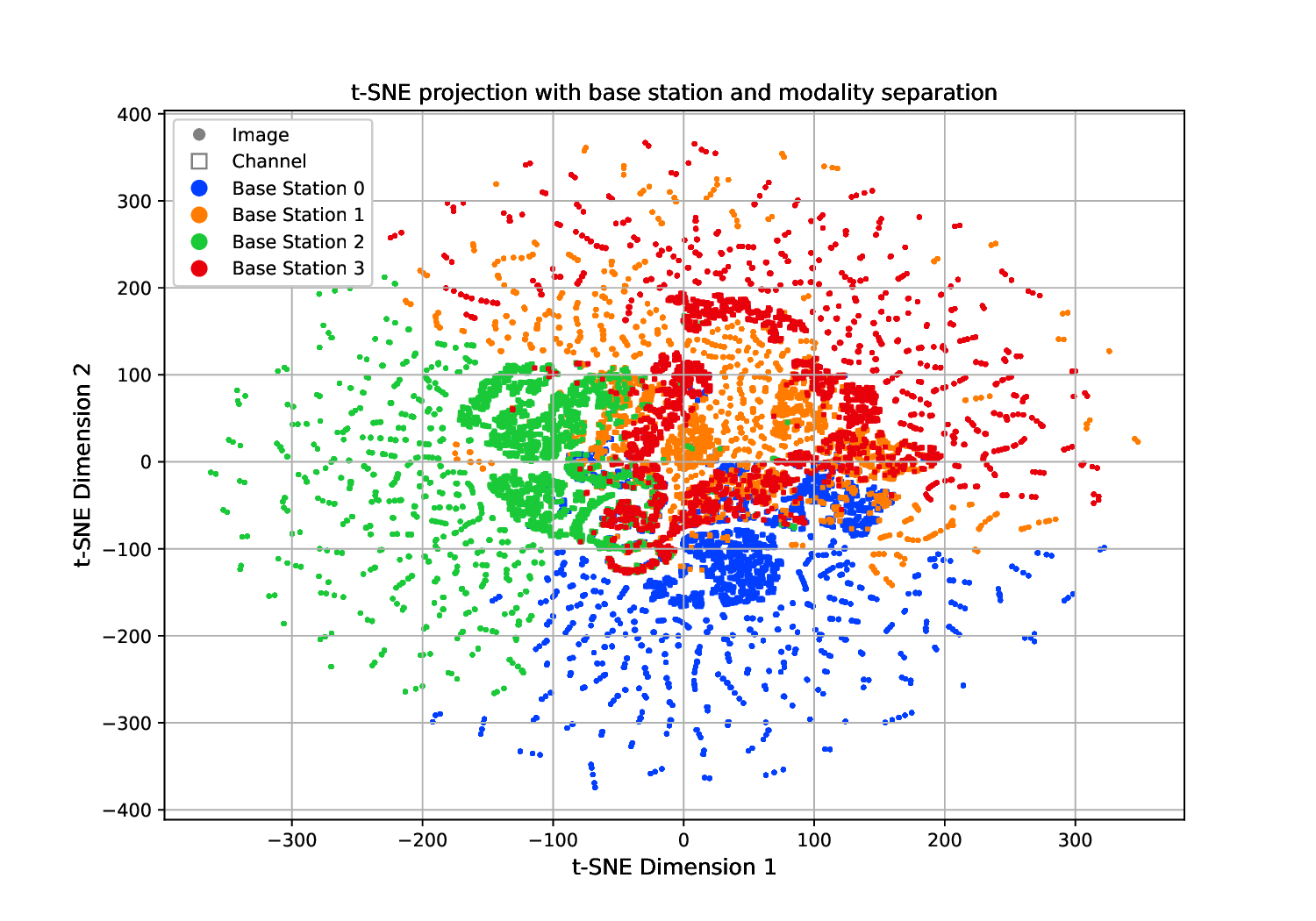}
        \caption{t-SNE of learned \gls{csi} and \gls{rgb} embeddings.}
        \label{fig:training:t-sne}
    \end{subfigure}
    \vspace{1em} 

    \begin{subfigure}{\columnwidth}
        \centering
        \includegraphics[width=\linewidth]{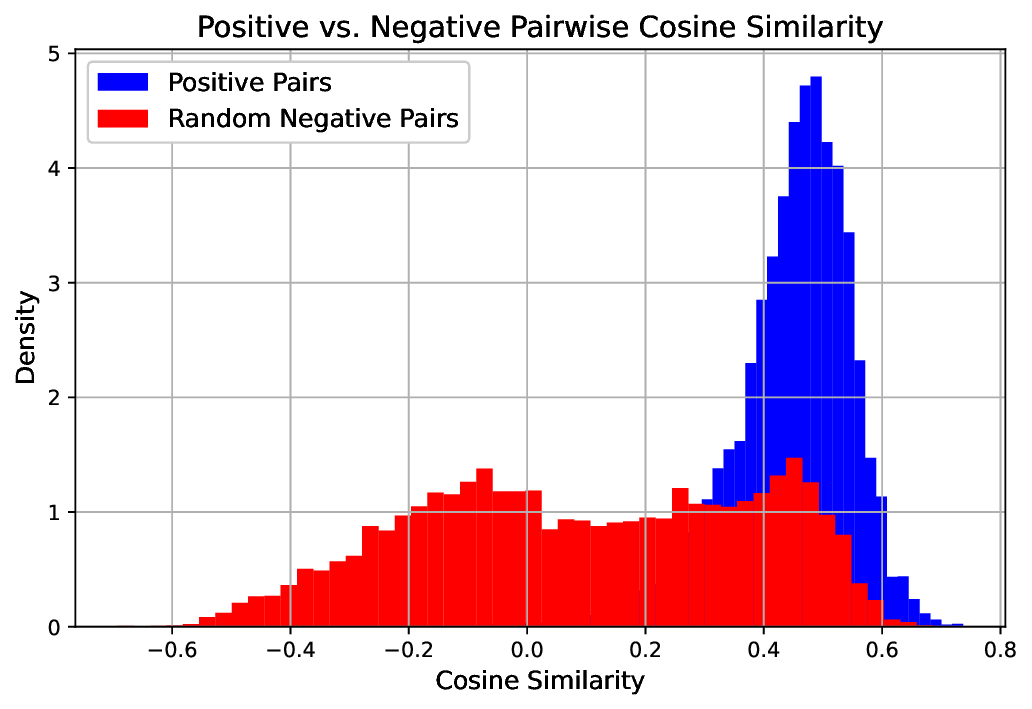}
        \caption{Cosine similarity between positive and negative pairs.}
        \label{fig:training:similarity}
    \end{subfigure}

    \caption{t-SNE visualization and cosine similarity distributions.}
    \label{fig:training_results}
\end{figure}
Fig. ~\ref{fig:training:t-sne} presents a two-dimensional \gls{tsne} projection of the learned multimodal embeddings, where \gls{tsne} is a nonlinear dimensionality reduction technique that visualizes high-dimensional data while preserving local similarity \cite{t_sne}. 
Distinct clusters corresponding to the four \glspl{bs} indicate that the proposed \gls{wmfm} effectively aligns image and channel embeddings of co-located samples within a shared spatial feature space.

Fig. ~\ref{fig:training:similarity} compares the cosine similarity distributions of positive and negative pairs in the learned joint embedding space. 
Positive pairs, formed from synchronized \gls{csi} and camera samples of the same \gls{bs}, exhibit high similarity values, while randomly mismatched pairs are centered near zero. 
This clear separation indicates that the proposed \gls{wmfm} effectively aligns cross-modal representations for true pairs while maintaining distinct embeddings for unrelated samples.

\begin{table}
\centering
\caption{Top-\(k\) \gls{bs} retrieval accuracy for channel and image modalities.}
\begin{tabular}{lcc}
\toprule
\textbf{} & \textbf{Channel Retrieval} & \textbf{Image Retrieval} \\
\midrule
Top-1 \gls{bs} Accuracy & 86.00\% & 77.62\% \\
Top-2 \gls{bs} Accuracy & 96.32\% & 77.62\% \\
Top-5 \gls{bs} Accuracy & 99.83\% & 77.70\% \\
\bottomrule
\end{tabular}
\label{tab:bs_retrieval}
\end{table}
Table~\ref{tab:bs_retrieval} reports the top-\(k\) \gls{bs} retrieval accuracy using the pretrained \gls{wmfm}. Two retrieval tasks were performed: channel retrieval (image-to-channel) and image retrieval (channel-to-image). In the channel retrieval setting, an image was provided as input, and the model retrieved the closest channel embedding based on cosine similarity. If the retrieved embedding corresponded to the correct \gls{bs}, the prediction was considered correct. The high top-1 and top-5 accuracy for both directions indicate strong cross-modal alignment, demonstrating that the learned latent representations are semantically consistent across the wireless and visual modalities.

\subsection{\gls{los}/\gls{nlos} Classification Results}

\begin{figure}
    \centering
    \begin{subfigure}[b]{0.48\columnwidth}
        \centering
        \includegraphics[width=\columnwidth]{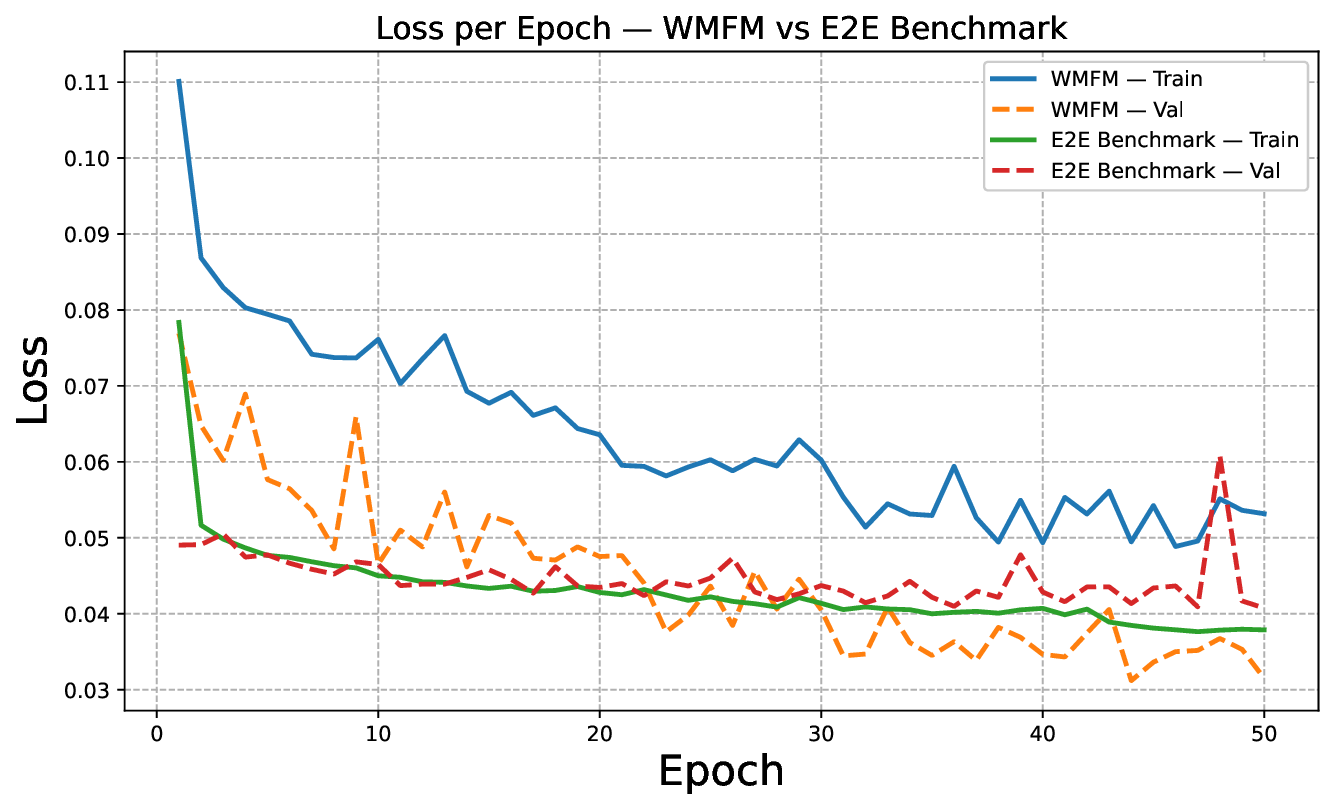}
        \caption{Training loss comparison.}
        \label{fig:los:loss}
    \end{subfigure}
    \hfill
    \begin{subfigure}[b]{0.48\columnwidth}
        \centering
        \includegraphics[width=\columnwidth]{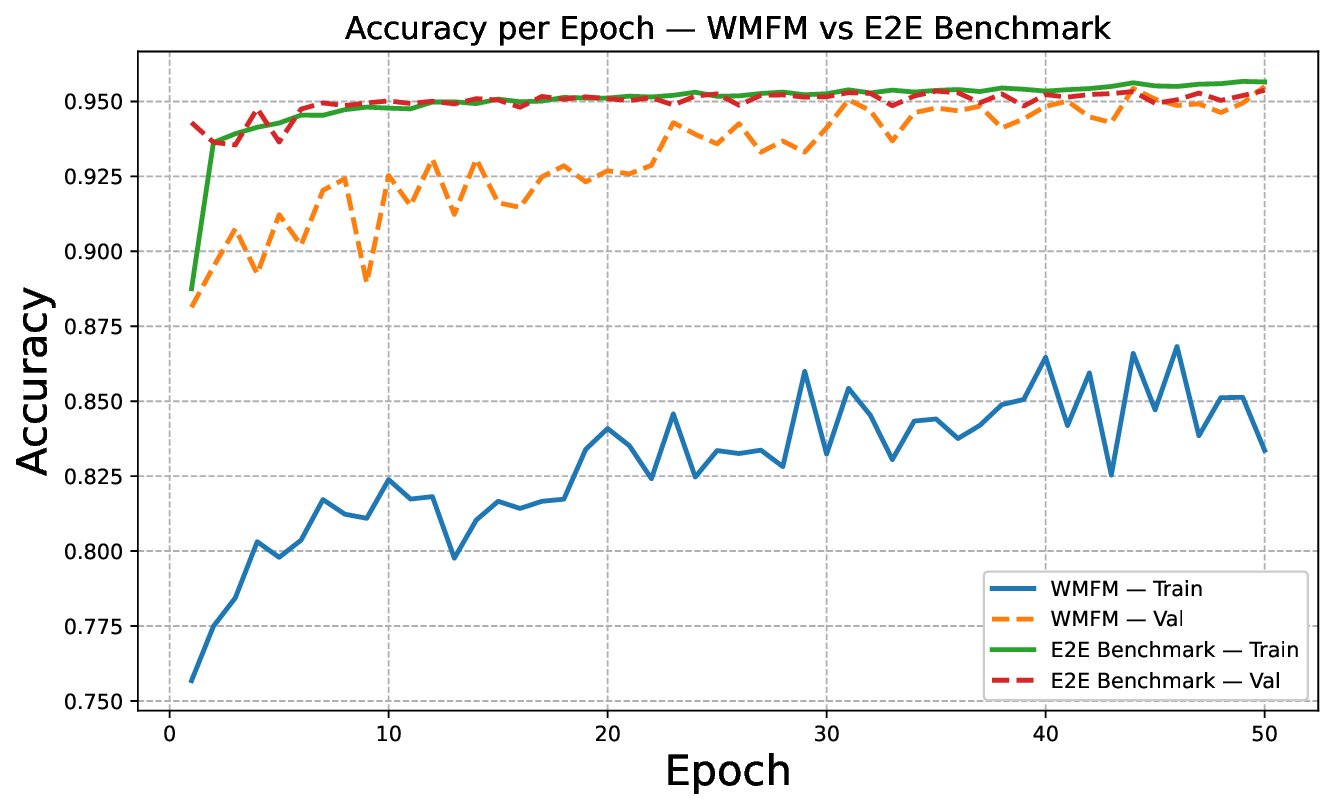}
        \caption{Accuracy.}
        \label{fig:los:acc}
    \end{subfigure}

    \vspace{1em}
    \begin{subfigure}[b]{0.48\columnwidth}
        \centering
        \includegraphics[width=\columnwidth]{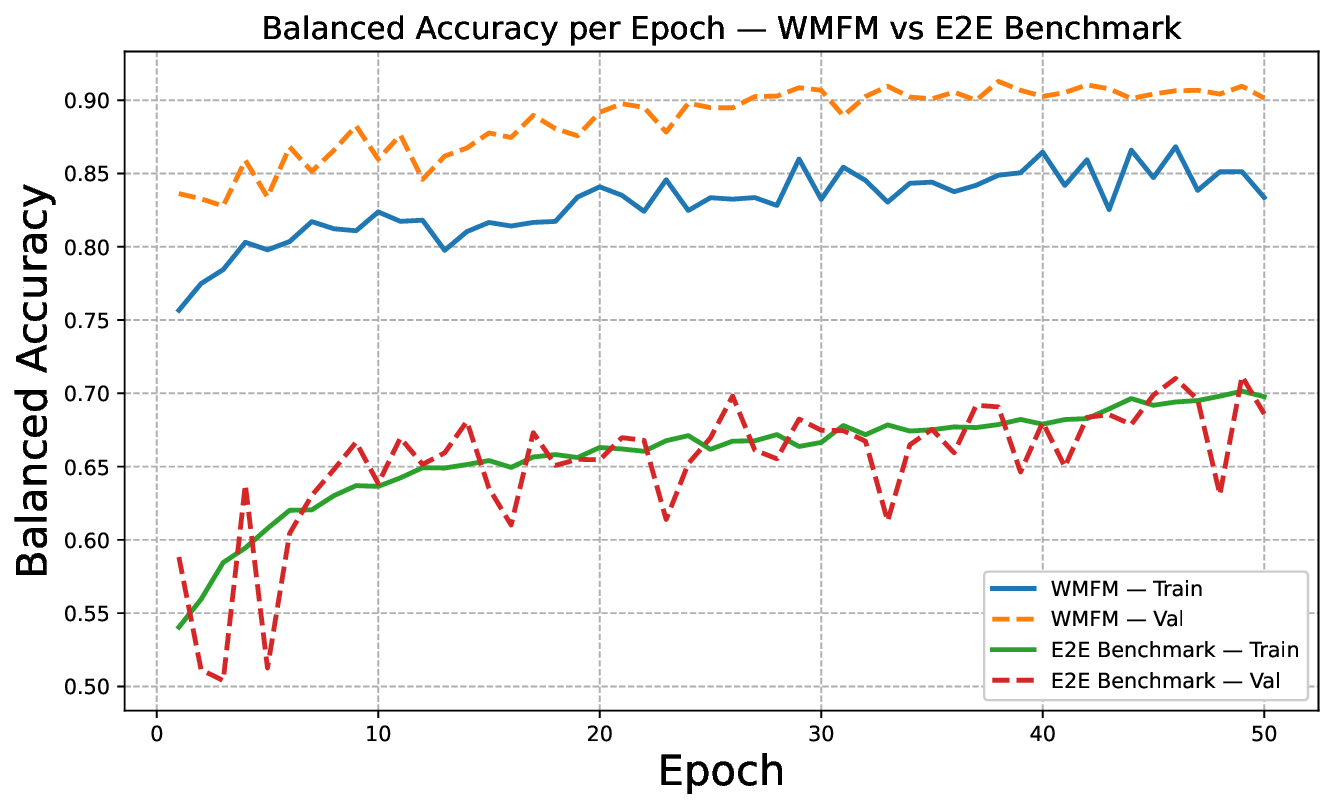}
        \caption{Balanced accuracy.}
        \label{fig:los:bal_acc}
    \end{subfigure}
    \hfill
    \begin{subfigure}[b]{0.48\columnwidth}
        \centering
        \includegraphics[width=\columnwidth]{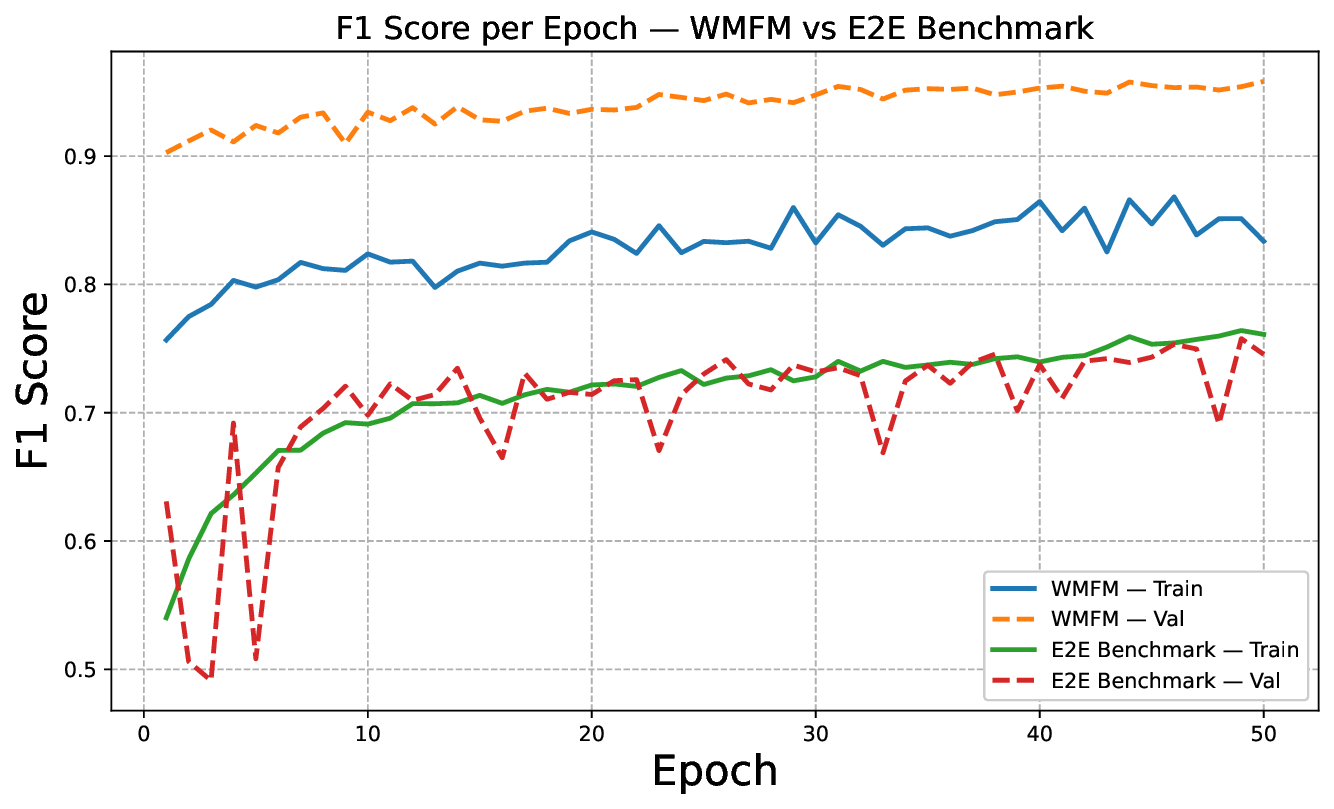}
        \caption{F1 score comparison.}
        \label{fig:los:f1}
    \end{subfigure}

    \caption{Comparison of the proposed model and benchmark across training and validation metrics.}
    \label{fig:los:plot}
\end{figure}

Fig. ~\ref{fig:los:plot} compares the evolution of training and validation metrics for the proposed \gls{wmfm} and the \gls{e2e} benchmark during \gls{los}/\gls{nlos} classification. 
Both models exhibit comparable training convergence. However, the \gls{wmfm} achieves consistently higher balanced accuracy and F1 scores, highlighting its superior capability to handle class imbalance and better distinguish minority \gls{nlos} samples.








\begin{table}
\centering
\caption{Performance comparison between \gls{wmfm}, \gls{e2e} classification benchmark, and \gls{csi} (single modality) task adaptation (Ablation Study) on \gls{los}/\gls{nlos} classification.}
\label{tab:los:performance_comparison}
\renewcommand{\arraystretch}{1.3} 
\setlength{\tabcolsep}{6pt}       

\begin{tabular}{|l|c|c|c|}
\hline
\textbf{Metric} & 
\textbf{\gls{wmfm}} & 
\textbf{\gls{e2e}} &
\textbf{\gls{csi} Task Adapt.} \\ 
\hline

Training time/epoch & 
3.16 s & 
277.4 s & 
2.52 s \\ 
\hline

Test Loss & 
\makecell{$0.0307$ \\ $\pm\,0.0004$} &
\makecell{$0.0390$ \\ $\pm\,0.0003$} &
\makecell{$0.0492$ \\ $\pm\,0.0004$} \\ 
\hline

Test Accuracy & 
\makecell{$95.69\%$ \\ $\pm\,0.26$} &
\makecell{$95.64\%$ \\ $\pm\,0.28$} &
\makecell{$92.40\%$ \\ $\pm\,0.35$} \\ 
\hline

Test Balanced Accuracy & 
\makecell{$90.65\%$ \\ $\pm\,0.85$} &
\makecell{$73.01\%$ \\ $\pm\,0.60$} &
\makecell{$82.27\%$ \\ $\pm\,1.31$} \\ 
\hline

\end{tabular}
\end{table}

Table~\ref{tab:los:performance_comparison} compares the proposed \gls{wmfm} with the \gls{e2e} for \gls{los}/\gls{nlos} classification. 
Although the overall accuracy appears similar due to the dominance of the \gls{los} class, it fails to capture the true performance disparity between models under class imbalance, highlighting the importance of using balanced accuracy metric.
The \gls{wmfm} achieves lower test loss and significantly higher balanced accuracy, indicating improved generalization and robustness to class imbalance. 
Moreover, it converges substantially faster, reducing training time per epoch by over two orders of magnitude compared to the \gls{e2e} baseline, owing to the frozen pretrained encoders and lightweight task head.

\begin{table}
\centering
\caption{Performance comparison between \gls{wmfm} (out of bracket) and \gls{e2e} benchmark (in bracket).}
\label{tab:los_nlos}
\renewcommand{\arraystretch}{1.2}
\setlength{\tabcolsep}{8pt}

\begin{tabular}{lccc}
\toprule
\textbf{Class} & \textbf{Precision} & \textbf{Recall} & \textbf{F1 Score} \\
\midrule
0 (\gls{nlos}) & 
0.62 (0.77) & 
0.85 (0.47) & 
0.72 (0.58) \\

1 (\gls{los}) & 
0.99 (0.96) & 
0.96 (0.99) & 
0.98 (0.98) \\
\bottomrule
\end{tabular}
\end{table}
Table~\ref{tab:los_nlos} reports the per-class precision, recall, and F1 score for the proposed \gls{wmfm} and the \gls{e2e} benchmark. 
Both models perform comparably on the dominant \gls{los} class; however, the \gls{wmfm} substantially improves recall and F1 score for the minority \gls{nlos} class, indicating enhanced sensitivity to challenging \gls{nlos} conditions. 
This demonstrates that the multimodal pretraining enables better discrimination of subtle propagation differences that the \gls{e2e} model fails to capture.

\subsubsection{Ablation Study}
To assess the contribution of multimodal fusion, we conduct an ablation study (Table \ref{tab:los:performance_comparison}) using only the channel encoder followed by a task-specific head, while keeping all other settings identical. 
This configuration leverages solely \gls{csi} features, as image-only training is infeasible due to multiple \glspl{ue} being visible within a single camera frame at a given time. 
As shown in Table~\ref{tab:los:performance_comparison}, integrating camera embeddings with channel embeddings substantially improves accuracy and balanced accuracy, demonstrating that visual context provides complementary information that enhances the discriminative power of the channel representations.

\subsubsection{Classification performance vs \% of training data used}
\begin{figure}
    \centering
    \includegraphics[width=1\linewidth]{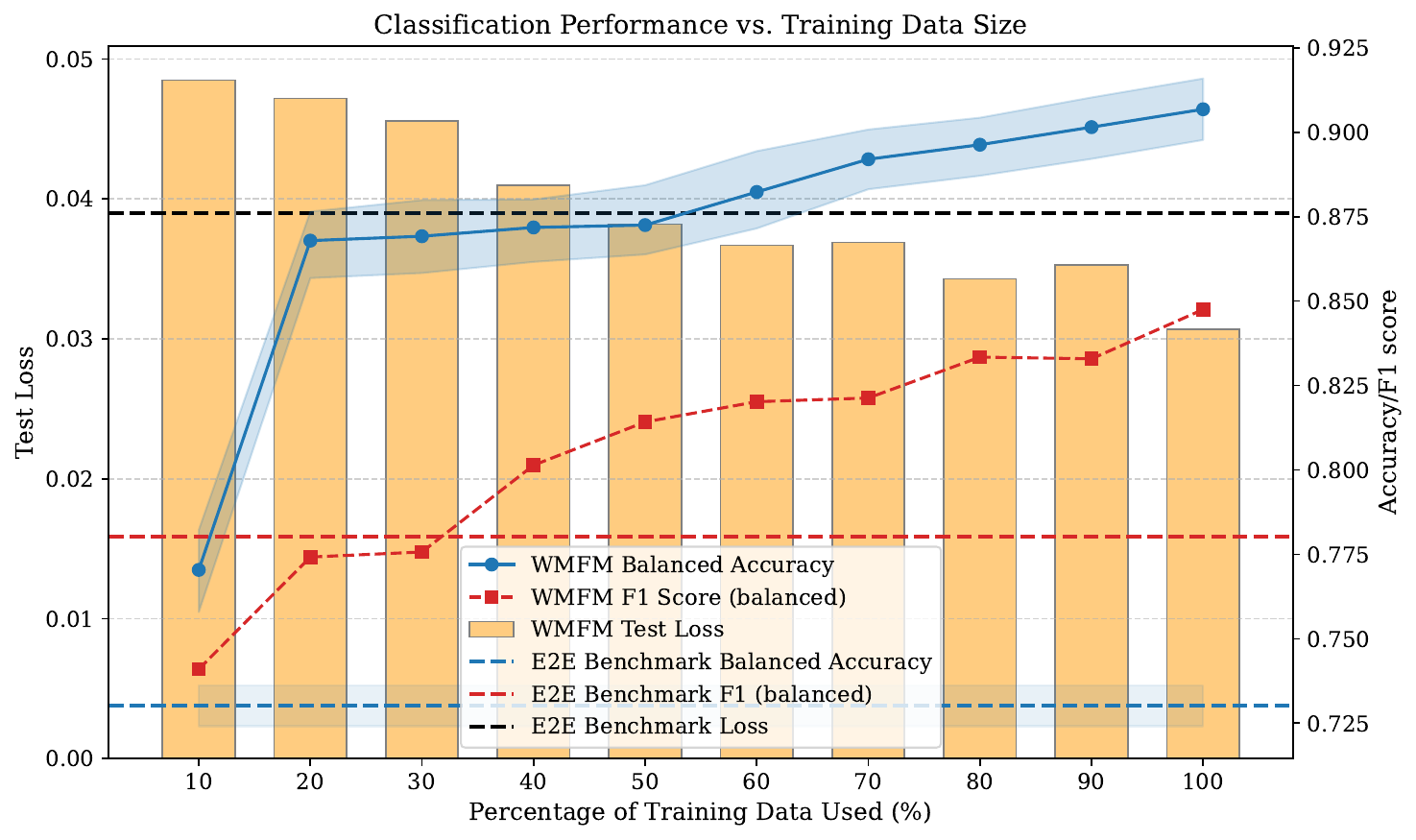}
    \caption{Comparison on test set between \gls{e2e} using 100\% of the training data, and \gls{wmfm} with different \% of training data for \gls{los}/\gls{nlos} classification.}
    \label{fig:los:dataused}
\end{figure}

Fig. ~\ref{fig:los:dataused} compares the performance of the proposed \gls{wmfm}, trained with varying fractions of the available data, against the \gls{e2e} benchmark trained on the full dataset. 
The \gls{wmfm} achieves comparable or superior balanced accuracy and F1 scores even when trained with a fraction of the data, while maintaining lower test loss. 
This performance gap highlights the data efficiency and generalization of \gls{wmfm}, which leverages semantically rich, modality-aligned representations from pretrained encoders, making downstream learning more effective and less reliant on large labeled datasets, unlike the more challenging \gls{e2e} training from scratch.
The \gls{e2e} model would likely require significantly more labeled data and longer training time to match or exceed the performance of our pretrained \gls{wmfm} approach.

\subsection{Localization Results}
\begin{figure}
    \centering
    \begin{subfigure}[b]{0.48\columnwidth}
        \centering
        \includegraphics[width=\columnwidth]{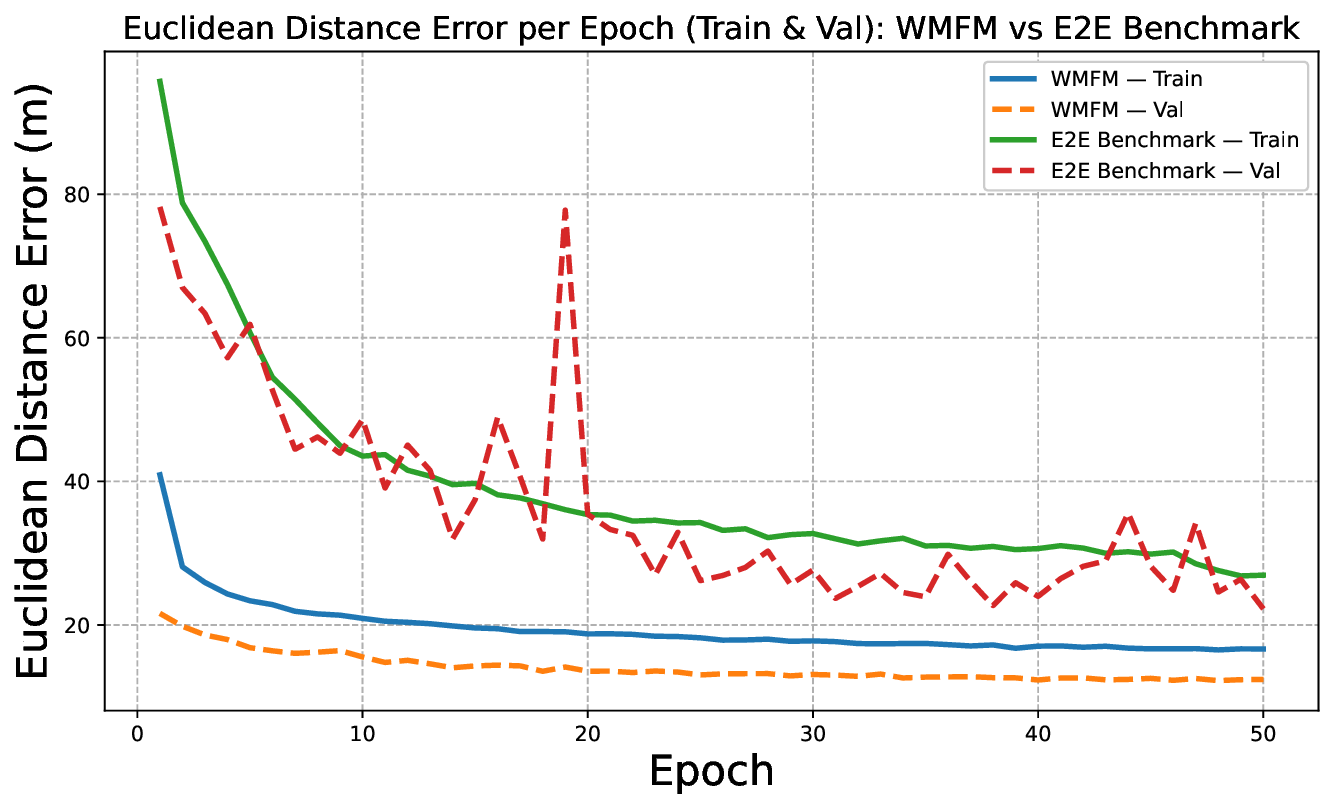}
        \caption{Euclidean Distance.}
        \label{fig:loc:euc}
    \end{subfigure}
    \hfill
    \begin{subfigure}[b]{0.48\columnwidth}
        \centering
        \includegraphics[width=\columnwidth]{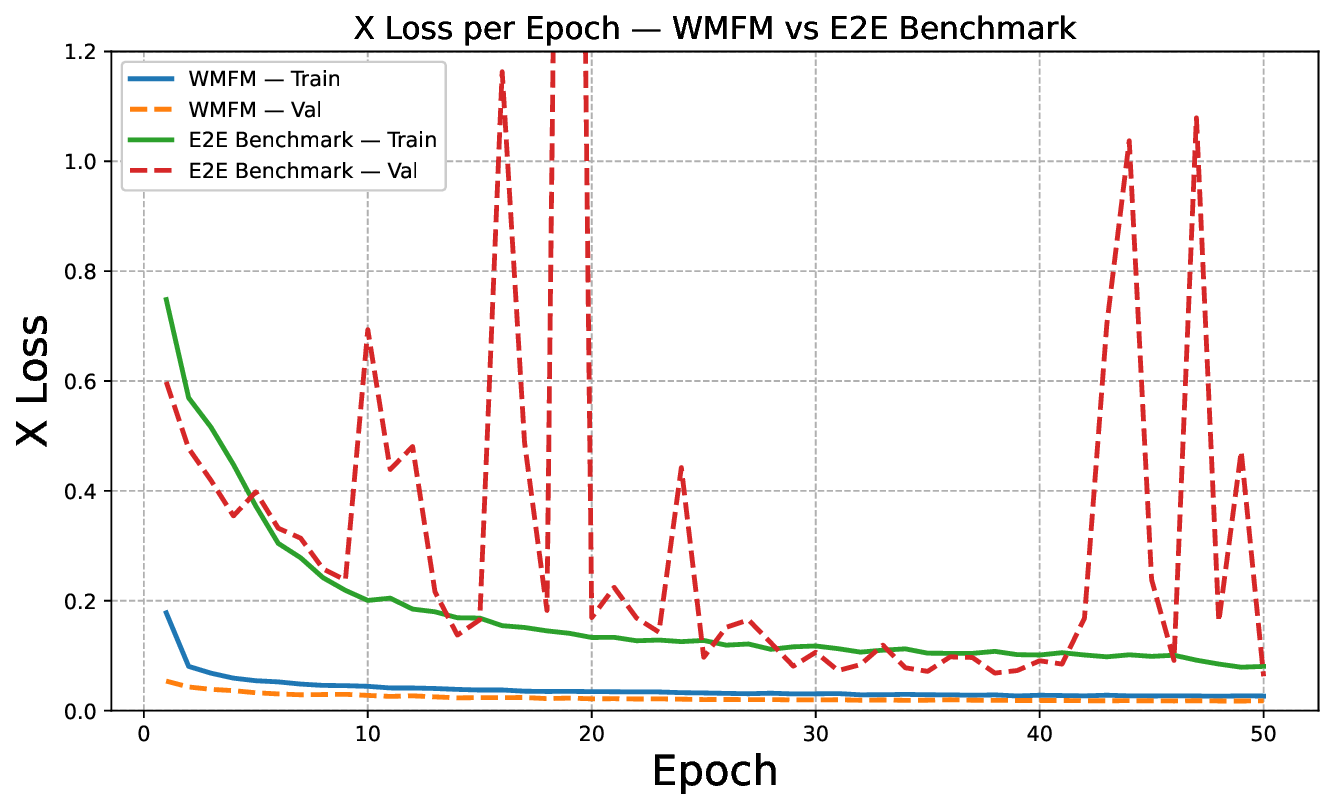}
        \caption{Loss for $x$-dimension.}
        \label{fig:loc:xloss}
    \end{subfigure}

    \vspace{1em}
    \begin{subfigure}[b]{0.48\columnwidth}
        \centering
        \includegraphics[width=\columnwidth]{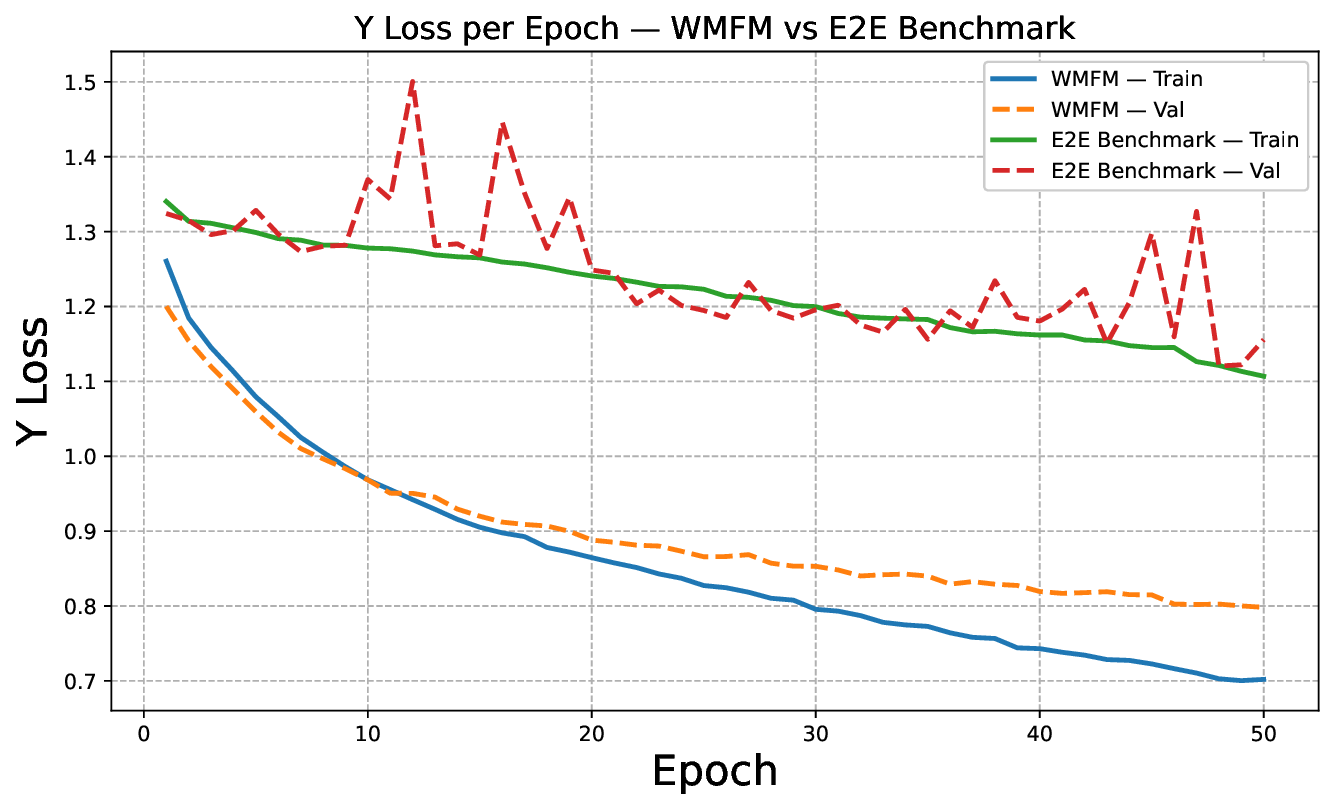}
        \caption{Loss for $y$-dimension.}
        \label{fig:loc:yloss}
    \end{subfigure}
    \hfill
    \begin{subfigure}[b]{0.48\columnwidth}
        \centering
        \includegraphics[width=\columnwidth]{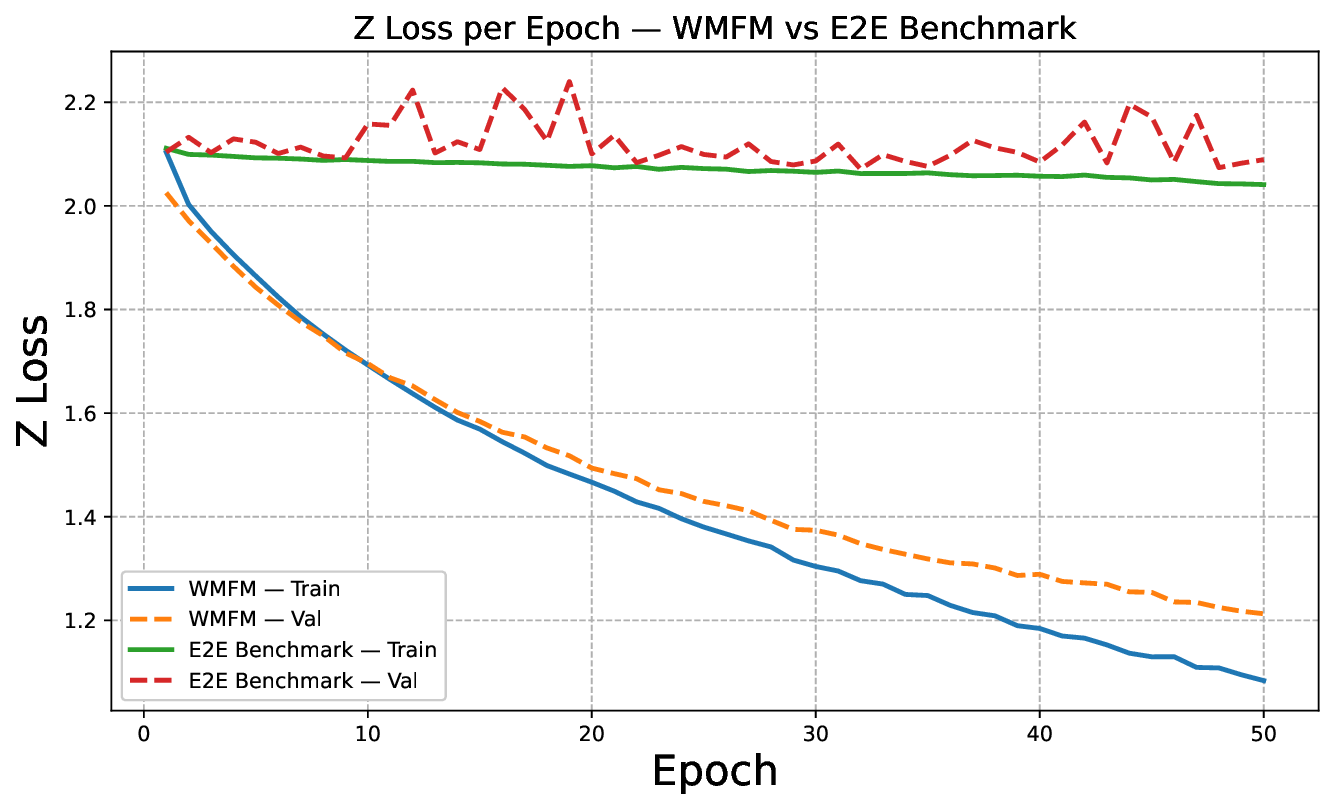}
        \caption{Loss for $z$-dimension.}
        \label{fig:loc:zloss}
    \end{subfigure}

    \caption{Comparison of the \gls{wmfm} and \gls{e2e} benchmark across training and validation metrics. 
    }
    \label{fig:loc:plot}
\end{figure}
Fig. ~\ref{fig:loc:plot} presents the training and validation trends for localization, comparing the proposed \gls{wmfm} with the \gls{e2e} benchmark. 
Across all subplots, the \gls{wmfm} demonstrates faster convergence, lower Euclidean distance, and consistently smaller coordinate-wise losses along the $x$, $y$, and $z$ dimensions. 
The distinct oscillations in the \gls{e2e} validation loss, particularly between epochs 10--20 and 40--50, suggest that the model struggles to settle into a robust minimum, likely due to the difficulty of learning complex spatial mappings from scratch.
Since the \gls{e2e} model learns from scratch, it is less efficient as it must infer task-relevant spatial features directly from raw data. With extended training and larger datasets, its performance could potentially approach that of the pretrained \gls{wmfm}.

\begin{table}
\centering
\caption{Performance comparison between \gls{wmfm}, \gls{e2e} classification benchmark, and \gls{csi} (single modality) task adaptation (Ablation Study) on localization.}
\label{tab:localization}
\renewcommand{\arraystretch}{1.3}
\setlength{\tabcolsep}{5pt} 

\begin{tabular}{|l|c|c|c|}
\hline
\textbf{Metric} &
\textbf{\gls{wmfm}} &
\textbf{\gls{e2e}} &
\textbf{\gls{csi} Task Adapt.} \\
\hline

Training time/epoch&
8.31 s &
701.3 s &
6.26 s \\
\hline

Test Avg. Euclidean Dist. &
\makecell{$12.37$ \\ $\pm\,0.17$ m} &
\makecell{$24.02$ \\ $\pm\,0.38$ m} &
\makecell{$27.14$ \\ $\pm\,0.41$ m} \\
\hline

X Loss (MSE) &
0.0162 &
0.0716 &
0.0884 \\
\hline

Y Loss (Cross-Entropy) &
0.806 &
1.160 &
1.047 \\
\hline

Z Loss (Cross-Entropy) &
1.204 &
2.075 &
1.915 \\
\hline

X Coordinate MAE &
\makecell{$11.52$ \\ $\pm\,0.18$ m} &
\makecell{$22.88$ \\ $\pm\,0.39$ m} &
\makecell{$26.27$ \\ $\pm\,0.42$ m} \\
\hline

Y Coordinate Accuracy &
\makecell{$0.6675$ \\ $\pm\,0.0053$} &
\makecell{$0.4994$ \\ $\pm\,0.0067$} &
\makecell{$0.5529$ \\ $\pm\,0.0066$} \\
\hline

Z Coordinate Accuracy &
\makecell{$0.5627$ \\ $\pm\,0.0034$} &
\makecell{$0.2262$ \\ $\pm\,0.0056$} &
\makecell{$0.2922$ \\ $\pm\,0.0061$} \\
\hline

\end{tabular}
\end{table}

Table~\ref{tab:localization} summarizes the localization performance of the proposed \gls{wmfm} and the \gls{e2e} benchmark. 
The \gls{wmfm} achieves substantially lower Euclidean and coordinate-wise errors, reflecting more accurate position estimation along all spatial dimensions. 
In addition, it converges nearly two orders of magnitude faster, highlighting the efficiency benefits of leveraging pretrained multimodal representations over learning localization mappings from scratch.

\begin{figure}
  \centering

  \begin{subfigure}{0.48\columnwidth}
    \centering
    \includegraphics[width=\linewidth]{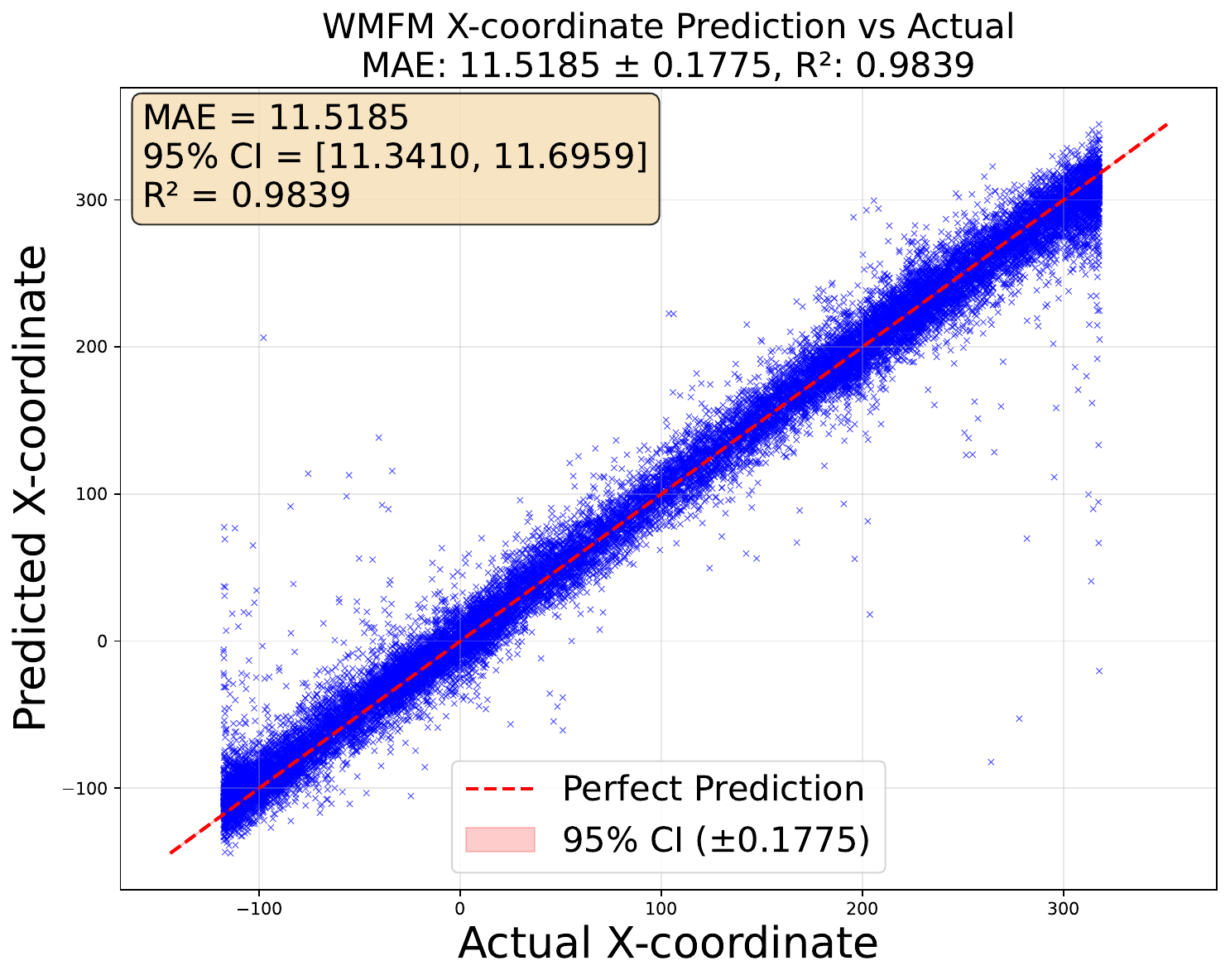}
    \caption{\gls{wmfm}}
    \label{fig:loc:ourx}
  \end{subfigure}
  \hfill
  \begin{subfigure}{0.48\columnwidth}
    \centering
    \includegraphics[width=\linewidth]{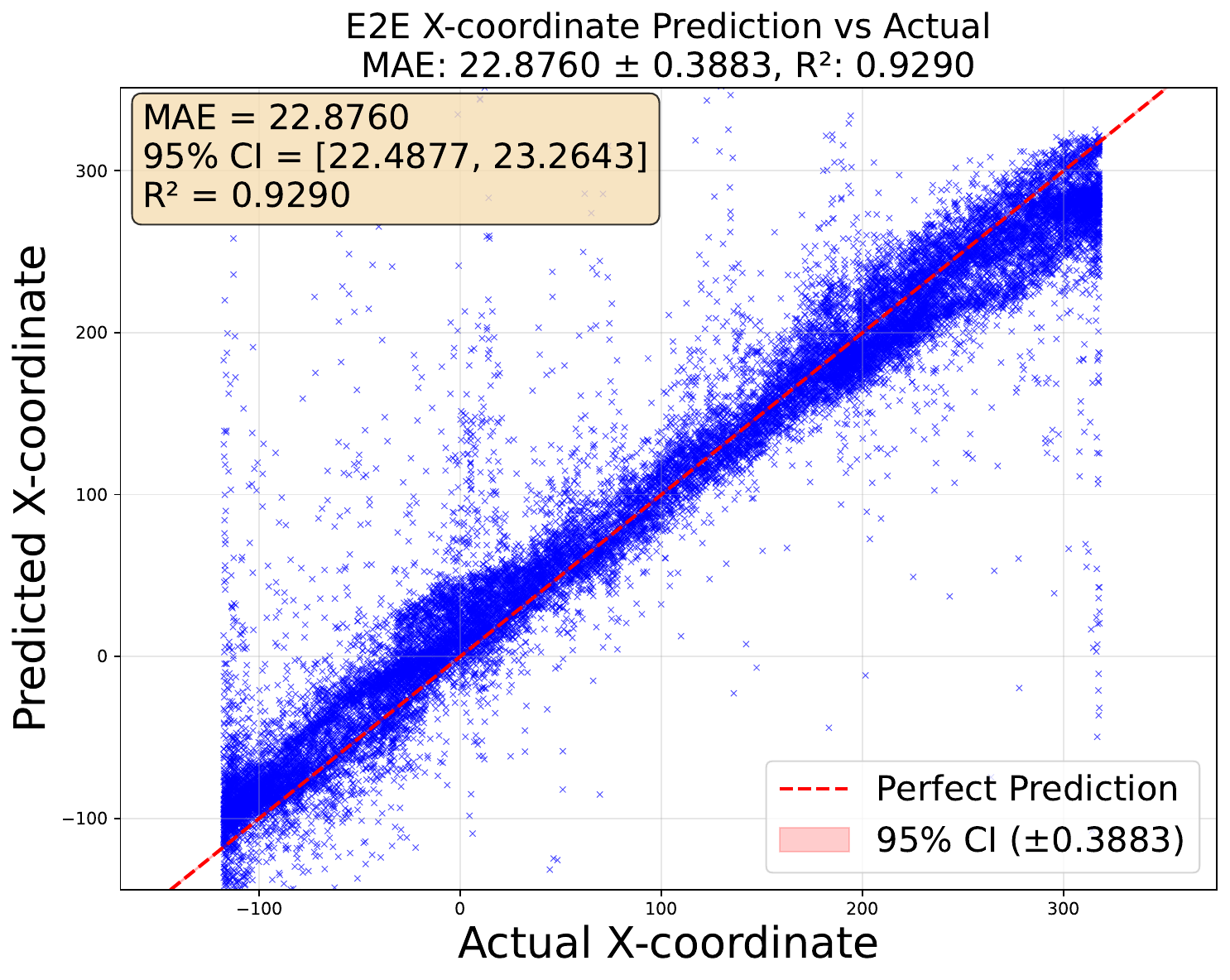}
    \caption{\gls{e2e} benchmark}
    \label{fig:loc:x_comp_bm}
  \end{subfigure}

  \caption{Actual and predicted x-coordinate comparison}
  \label{fig:loc:x_comp}
\end{figure}
Fig.~\ref{fig:loc:x_comp} compares the actual versus predicted \( x \)-coordinates using the proposed \gls{wmfm} model and the \gls{e2e} benchmark. The scatter plots illustrate the correlation between predicted and ground-truth values, with the red dashed line representing perfect prediction. \gls{wmfm} achieves significantly higher accuracy, with a lower mean absolute error, and a higher coefficient of determination, indicating tight alignment between predicted and true values. In contrast, the \gls{e2e} shows greater prediction variance and reduced R2 score. 

\subsubsection{Ablation Study}
To evaluate the contribution of multimodal fusion, an ablation experiment was conducted using only the channel encoder followed by a localization head. 
As shown in Table \ref{tab:localization}, this single-modality configuration performs notably worse than the full \gls{wmfm}, with higher Euclidean distance and reduced coordinate accuracy, particularly along the $y$- and $z$-axes. 
By fusing camera embeddings with channel features, the model can associate visual structures with corresponding propagation patterns, leading to more precise position estimation and improved spatial consistency.

\subsubsection{Classification performance vs \% of training data used}

\begin{figure}
    \centering
    \includegraphics[width=1\linewidth]{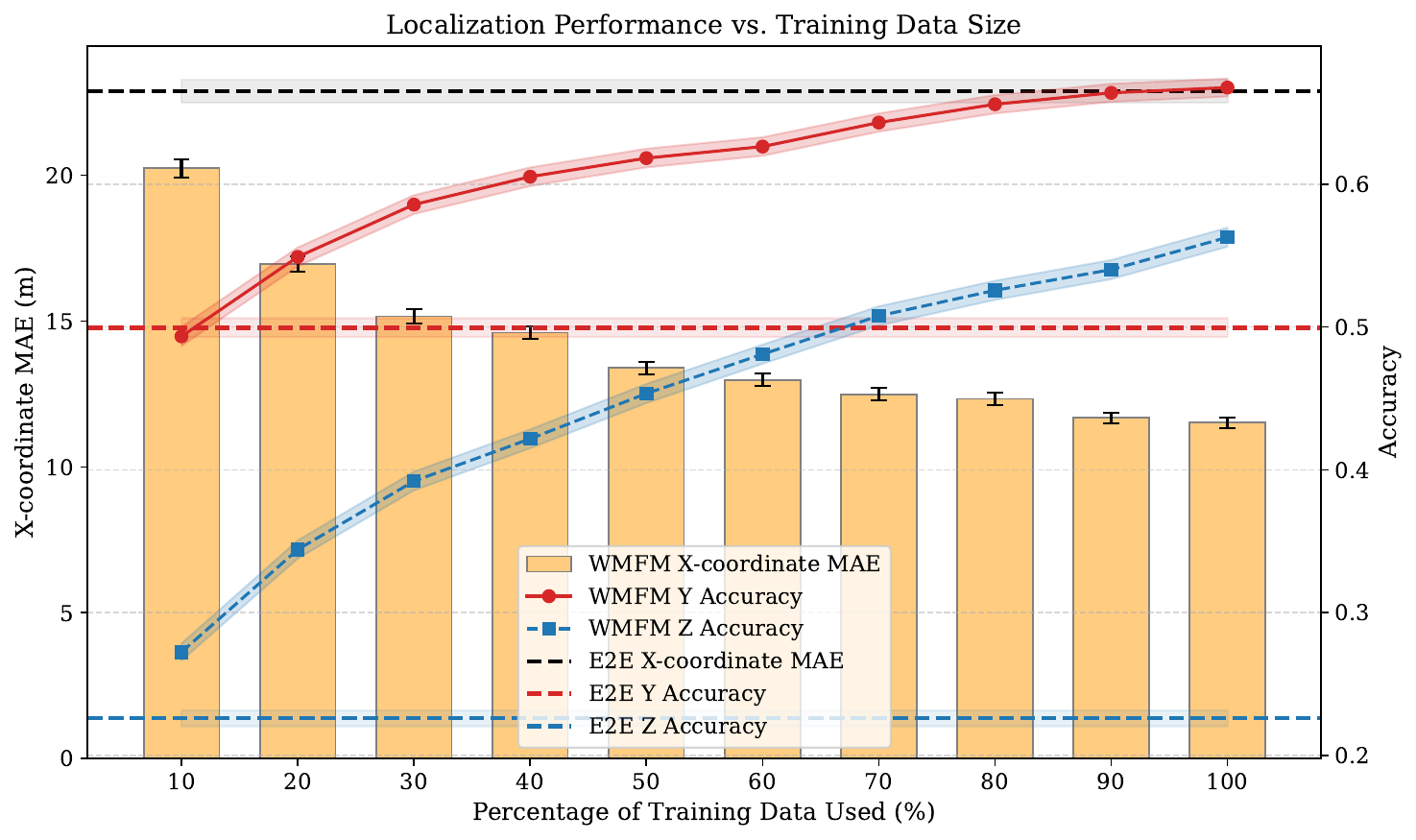}
    \caption{Comparison on test set between \gls{e2e} using 100\% of the training data, and \gls{wmfm} with different \% of training data for localization.}
    \label{fig:loc:data}
\end{figure}

Fig. ~\ref{fig:loc:data} compares the localization performance of the proposed \gls{wmfm}, trained with varying portions of the dataset, against the \gls{e2e} benchmark trained on the full data. 
The \gls{wmfm} maintains competitive or superior performance across all metrics, achieving low $x$-coordinate error and high $y$- and $z$-axis classification accuracy even with limited data. 
This demonstrates the strong data efficiency of the pretrained multimodal representations, which enable accurate spatial estimation and generalization under reduced supervision.

\subsubsection{Effect of noise on performance}

\begin{figure}
    \centering
    \includegraphics[width=1\linewidth]{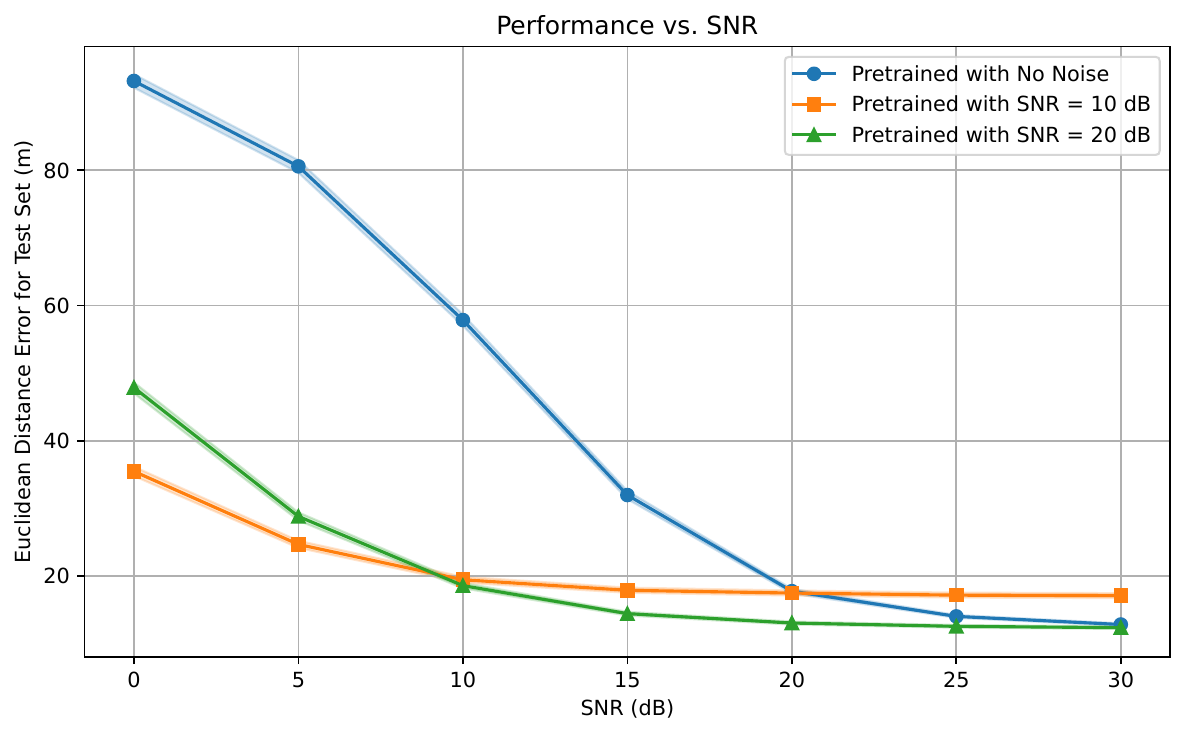}
    \caption{Localization error versus \gls{snr} 
    } \vspace{-2ex}
    \label{fig:loc:noise}
\end{figure}
Fig. ~\ref{fig:loc:noise} shows the impact of channel noise on localization performance, evaluated as the Euclidean distance error under varying \gls{snr} levels. 
While the model trained without noise performs best under clean conditions, it degrades rapidly as \gls{snr} decreases, revealing limited robustness to channel corruption. 
In contrast, models pretrained with moderate noise (e.g., 20~dB \gls{snr}) maintain low localization error across all test conditions, demonstrating improved resilience and generalization. 
These results confirm that controlled noise injection during training enhances the robustness of the learned multimodal representations without sacrificing high-\gls{snr} accuracy.
Moreover, the model pretrained with a lower \gls{snr} of 10~dB converges to a higher localization error. This degradation can be attributed to excessive noise overwhelming the \gls{csi} modality during pretraining, which limits the model's ability to learn informative and transferable features across modalities.

\section{Conclusion and Future Work} \label{section:conclusion}

In this paper, we introduced the contrastive learning based Wireless Multimodal Foundation Model, a novel framework that unifies wireless channel coefficients and visual imagery for multimodal representation learning in next-generation communication systems. The proposed model is pretrained using a contrastive self-supervised learning paradigm that aligns the embeddings of heterogeneous modalities without requiring labeled data. The pretrained encoders are subsequently employed as frozen feature extractors, enabling efficient adaptation to downstream tasks such as localization and \gls{los}/\gls{nlos} classification through lightweight task-specific heads.

Comprehensive experiments on the DeepVerse6G dataset demonstrated that the proposed \gls{wmfm} significantly outperforms both unimodal and \gls{e2e} baselines, achieving a 17\% improvement in \gls{los}/\gls{nlos} classification balanced accuracy and a 48.5\% reduction in localization error, while reducing training time by up to 90-fold. These results underscore the effectiveness of leveraging pretrained multimodal encoders to enhance generalization and data efficiency in integrated sensing and communication (\gls{isac}) systems. Furthermore, the theoretical analysis provided insights into the role of contrastive learning in maximizing mutual information and promoting alignment and uniformity across modalities.

For future work, we plan to extend the \gls{wmfm} framework by incorporating additional sensing modalities such as radar and LiDAR to enhance cross-modal representation learning. 
We also aim to explore fine-tuning strategies using diffusion or reinforcement learning to improve robustness under dynamic channel conditions. 
Finally, integrating the model into agentic or autonomous network management frameworks represents a promising direction toward intelligent and adaptive 6G systems.

\vspace{-2ex}
\section*{Acknowledgment}
This work has been supported by MITACS and Ericsson Canada. The authors also wish to honor the memory of their co-author, Dr.~Han~Zhang, whose remarkable contributions, dedication, and spirit continue to inspire this research.

\balance
\bibliographystyle{IEEEtran} 
\bibliography{journal_contrastive}
\balance \balance


\end{document}